\pdfoutput=1
\documentclass[10pt,twocolumn,amsmath,amssymb,longbibliography,nofootinbib,showkeys,eprint]{revtex4-2}

\usepackage[activate={true,nocompatibility},final,tracking=true,kerning=true,spacing=true]{microtype}
\usepackage{epigraph}
\usepackage{graphicx}
\usepackage{epstopdf}
\usepackage{braket}
\usepackage{bm}
\usepackage{numprint}
\usepackage{float}
\usepackage[T2A,T1]{fontenc}
\usepackage[utf8]{inputenc}
\usepackage{comment}
\usepackage{hhline}
\usepackage{xfrac}
\usepackage{amsthm}
\usepackage{makecell}
\usepackage{dsfont}
\usepackage{calc}
\usepackage{seqsplit}
\usepackage{circuitikz}
\usepackage{hyperref}
\hypersetup{
    colorlinks=true
}
\usepackage{orcidlink}

\begin{document}
% $Id: ExecutionFlow.tex,v 1.224 2026/04/02 21:56:38 mal Exp $
\preprint{V.M.}
\title{Trade Execution Flow as the Underlying Source of Market Dynamics}

\author{Mikhail Gennadievich \surname{Belov}}
\email{mikhail.belov@tafs.pro}
\affiliation{Lomonosov Moscow State University,  Faculty of Mechanics and Mathematics,
   GSP-1,  Moscow, Vorob'evy Gory, 119991, Russia}
\affiliation{Autretech Group, Skolkovo Innovation Center, Nobel Street, Building 7, Moscow, 121205, Russia}

\author{Victor Victorovich \surname{Dubov}}
\email{dubov@spbstu.ru}
\affiliation{Peter the Great St. Petersburg Polytechnic University, 195251, Russia}

\author{Vadim Konstantinovich \surname{Ivanov}\,\orcidlink{0000-0002-3584-4583}}
\email{ivvadim@rambler.ru}
\affiliation{Peter the Great St. Petersburg Polytechnic University, 195251, Russia}

\author{Alexander Yurievich \surname{Maslov}\,\orcidlink{0009-0005-6296-5988}}
\email{maslov.ton@mail.ioffe.ru}
\affiliation{Ioffe Institute, Politekhnicheskaya 26, St Petersburg, 194021, Russia}

\author{Olga Vladimirovna \surname{Proshina}\,\orcidlink{0000-0003-0087-5838}}
\email{proshina.ton@mail.ioffe.ru}
\affiliation{Ioffe Institute, Politekhnicheskaya 26, St Petersburg, 194021, Russia}

\author{Vladislav Gennadievich \surname{Malyshkin}\,\orcidlink{0000-0003-0429-3456}} 
\email{malyshki@ton.ioffe.ru}
\affiliation{Ioffe Institute, Politekhnicheskaya 26, St Petersburg, 194021, Russia}

\date{June, 6, 2024}

\begin{abstract}
\begin{verbatim}
$Id: ExecutionFlow.tex,v 1.224 2026/04/02 21:56:38 mal Exp $
\end{verbatim}
In this work, we demonstrate experimentally that the execution flow, $I = dV/dt$, is the fundamental driving force of market dynamics.
We develop a numerical framework to calculate execution flow from the data using the Radon-Nikodym derivative. A notable feature of this approach is its ability to automatically determine thresholds that can serve as actionable triggers.
The technique also determines the characteristic time scale directly from the corresponding eigenproblem.
The methodology has been validated on actual market data to support these findings.
Additionally, we introduce a framework based on the Christoffel function spectrum, which is invariant under arbitrary
non-degenerate linear transformations of input attributes and offers an alternative
to traditional principal component analysis (PCA), which is limited to unitary invariance.
\end{abstract}
\maketitle

\newpage

\section{\label{intro}Introduction}
Modern financial markets display complex dynamics arising from internal and external factors,
and from stochastic (or deterministic) processes not linked to any identifiable cause.
Since Aristotle \cite{polanyi1957aristotle}, this has been a fascinating topic of study, particularly price formation.
Price formation driven by market microstructure is the focus of this paper.
Most interestingly, the t\^{a}tonnement process \cite{walras2013elementsBOOK},
used as a means to observe supply and demand curves, misses the entire aspect of market dynamics \cite{donier2016walrasPAPER}.

Modern financial markets generate a diverse array of information, including prices, execution volumes across different time scales, limit order book (LOB) data from exchanges, corporate financial reports, sovereign economic indicators, central bank actions, and more.
The accessibility, structure, time scale, and impact of this information on market behavior vary significantly.

In \cite{ArxivMalyshkinMuse}, we formulated the ultimate market dynamics problem:
to find evidence of the existence (or proof of the non-existence)
of an automated trading machine that consistently generates positive P\&L (Profit and Loss) in a free market as an autonomous agent.
In \cite{MalMuseScalp}, we formulated the problem in weak and strong forms:
\begin{itemize}
\item Weak form: Whether such an automated trading machine can exist at all using only legally available data.
(It can certainly exist in an illegal form—for example, when a brokerage uses client order flow information to
\href{https://en.wikipedia.org/wiki/Front_running}{frontrun}
their own clients. Such strategies typically rely on proprietary information about clients' future supply-demand imbalances and on subsequent monetization of this information.)

\item Strong form: Whether such an automated trading machine can exist based solely on transaction sequences ---
for instance, the historical time series of market observation triples: (time, execution price, shares traded).
In this information, supply and demand are matched for every observation: at time $t$, trader $A$ sold $v$
shares of a security at price $P$ to trader $B$ and received $vP$ dollars.
Such a strategy can utilize only information about volume and execution flows.
\end{itemize}

In this paper, we focus on determining information about the future solely from a sequence of past execution triples: 
(time,execution price,shares traded).
The main observations of our previous works \cite{2015arXiv151005510G,2016arXiv160204423G}
is that it is the share execution flow  $I=dV/dt$,
rather than the share trading volume $V$,
that drives the market \cite{2016arXiv160204423G} (see Fig. \ref{ntPlotChart} below: the asset price exhibits singularities at high $I$,
whereas no price singularity occurs at the maximal volume price -- the median of the price-volume distribution).
In other words, it is the execution flow $I=dV/dt$, not the traded volume, that drives the market.
This perspective differs significantly from the commonly studied\cite{bucci2019crossover} concept
of \href{https://en.wikipedia.org/wiki/Market_impact\#Market_impact_cost}{market impact}.
The situation is analogous to the difference between Newtonian and Aristotelian dynamics:
force causes acceleration vs force causes velocity.
This difference is clearly demonstrated in Fig. \ref{ntPlotChart} below.

In this paper, we investigate market microstructure using trading data with sub-microsecond temporal resolution.
Previous research initiatives -- beginning with the Penn-Lehman Automated Trading (PLAT) project \cite{kearns2003penn}
and followed by others \cite{lebaron2006agent,chakole2021q} among many others --
have explored the performance characteristics of a variety of automated trading systems.
While our group has previously conducted high-frequency trading (HFT) on NASDAQ,
the present study focuses primarily on market microstructure analysis,
emphasizing execution flow as the fundamental driving mechanism of market dynamics.
The principal contributions of this work are as follows:
\begin{enumerate}
\item
Development of a fast and numerically stable method for moment calculation (Section \ref{momentsCalculation}).
\item
Application of this method to real exchange data (Section \ref{FinData}).
\item
Development of an execution flow estimation methodology (Section \ref{ExecutionFlowCalculation})
and experimental evidence linking execution flow singularities to price singularities.
The most important result is the automatic determination of the characteristic time scale from the corresponding eigenproblem.
\item
Derivation of a procedure for converting execution flow fluctuations into probabilistic forecasts of P\&L (Sections \ref{PnLExpressions} and \ref{impactF}).
\item
Empirical comparison of the derived directional information with observed market behavior (Section \ref{directionalInfo}).
\end{enumerate}
Additionally, we propose a framework based on the Christoffel function spectrum
for determining probability contribution components (Appendix \ref{ChristoffelSpectrum}),
which is invariant under arbitrary non-degenerate transformations of input attributes.
This invariance property provides a significant advantage over conventional principal component analysis (PCA),
which is limited to unitary invariance.

This paper is accompanied by a software which
\href{http://www.ioffe.ru/LNEPS/malyshkin/code_polynomials_quadratures.zip}{is available}
from Ref. \cite{polynomialcode};
all references to code in the paper correspond to this software.
A detailed description of its usage is provided in Appendix \ref{SoftwareDescription}.
All the exchange data used in this paper is available for download from the links in \cite{FinancialData};
also, see the exchange sites \cite{itchfeed,NYSEtaq} for more recent data.

\section{\label{momentsCalculation}Moment Calculation from Empirical Samples}
Having established the role of the execution flow  $I=dV/dt$,
we now formulate a method for its calculation.
For a given time series  $t_l,f_l$, we introduce the moments
$\Braket{Q_j f}$
calculated as
\begin{align}
\Braket{Q_j f}&=\int\limits_{-\infty}^{t_{now}} Q_j(x(t)) f(t) \omega(t)dt \nonumber \\
&=
\sum\limits_{l} Q_j(x(t_l)) f_l\, \omega^{(l)} (t_l -t_{l-1})
\label{momentsDef}
\end{align}
this sums the terms from the past till $t_{now}$.
Here, $x(t)$ is a monotonic function; in this paper, we use either
$x=(t-t_{now})/\tau$ or $x=\exp((t-t_{now})/\tau)$.
The function $\omega(t)$
is a decaying weight; in this paper, we consider only an exponential decay, 
$\omega=\exp((t-t_{now})/\tau)$.
The function $Q_j(x)$
is a polynomial of degree $j$.
One can simply use, for example, $Q_j(x)=x^j$,
but it is convenient to employ an arbitrary basis to improve numerical stability.
In this paper, we often use the basis of shifted Legendre polynomials:
$Q_j(x(t))=P_j\left(2\exp((t-t_{now})/\tau)-1\right)$,
where $P_j(x)$ 
denotes the Legendre polynomial of degree $j$.

Equation (\ref{momentsDef}) is simply an exponential moving average of
$Q_j(x(t))f(t)$.
For example, a regular moving average price $P_{ma}$
and moving standard deviation $\sigma_{ma}$,
calculated from a sequence $(t_l,P_l)$, is given by
\begin{align}
P_{ma}(t_{now})&=\frac{\Braket{Q_0 P}}{\Braket{Q_0}} \label{pMovaverage} \\
\sigma^2_{ma}(t_{now})&=\frac{\Braket{Q_0 P^2}}{\Braket{Q_0}} - P^2_{ma}(t_{now}) \label{sigmaMovaverage}
\end{align}
Equation (\ref{momentsDef}) maps a long sequence of past observations $t_l,f_l$ to $n$ moments $\Braket{Q_j f}$,
with $j=0\dots n-1$. The index $j$ captures contributions from different time scales.
If one chooses $Q_j(x(t)) = \exp(i\, jt /\tau)$ and $\omega = 1$,
the moments $\Braket{Q_j f}$ correspond essentially to Fourier amplitudes.
In this work, we adopt a decaying weight and an arbitrary basis $Q_j(x)$
to improve numerical stability and better capture the dynamics of interest.

Given a sequence of  (time, execution price, shares traded) as $(t_l, P_l, dV_l)$\footnote{
For convenience, we define $dV_l = V_l - V_{l-1}$ as the number of shares traded at $t_l$,
where $V_l$ denotes the total volume traded at or before $t_l$.
}
Consider all possible moments that can be calculated from such sequences.
They essentially differ only in the choice of integration variable in (\ref{momentsDef});
instead of $t_l - t_{l-1}$, one can use $P_l - P_{l-1}$ or $ V_l - V_{l-1}=dV_l$.
Consider, for example, the execution flow (trade execution rate):
\begin{align}
I &= \frac{dV}{dt} \approx \frac{V_l - V_{l-1}}{t_l - t_{l-1}}
\label{IclassicDef}
\end{align}
The choice of integration variable allows us to calculate different rates.
We now list all the moments that can be calculated by direct sampling using the definition (\ref{momentsDef})
with the following measures:
\begin{subequations}
\label{measuresList}
\begin{align}
dt=t_l-t_{l-1}&\quad \text{for $\Braket{P^k Q_j }$} \label{dtmuSample} \\
dP=P_l-P_{l-1}&\quad \text{for $\Braket{P^k Q_j \frac{dP}{dt}}$, $\Braket{P^k Q_j V \frac{dP}{dt}}$} \label{dPmuSample}  \\
dV=V_l-V_{l-1}&\quad \text{ for $\Braket{P^k Q_j \frac{dV}{dt}}$} \label{dVmuSample}  \\
d\varpi=t_l-t^{\mathrm{in}}_l&\quad \text{ for $\Braket{P^k Q_j \frac{d\varpi}{dt}}$} \label{dvpmuSample} 
\end{align}
\end{subequations}
Other moments, such as $\Braket{P^k Q_j \frac{dPV}{dt}}$, can be obtained from (\ref{measuresList}) using integration by parts.
The measure $d\varpi$ (\ref{dvpmuSample}) represents, for an execution at $t_l$,
the time the initial limit order (which arrived at $t^{\mathrm{in}}_l$) spent in the
LOB before being matched by another order at time $t_l$ to produce an execution.
This measure allows for the construction of the true execution rate
$\widetilde{I}$ (\ref{imodified}), considered in Appendix \ref{DirAssym} below.
Note that while the integrals
$\int dt$, $\int dP$, and $\int dV$
have clear meanings of elapsed time, price change, and total volume traded, the integral  $\int d\varpi$
has no such total change meaning, i.e., it is not an ``extensive'' function.
This may prevent the usage of $d\varpi$
as an integration measure to calculate $\widetilde{I}$.
A na\"{\i}ve option is to modify (\ref{dvpmuSample}) to obtain a measure that behaves as an extensive function;
for example, a volume-weighted time in the LOB
\begin{align}
d\widetilde{\varpi}=\left(t_l-t^{\mathrm{in}}_l\right)\left(V_l-V_{l-1}\right)  \label{dvpmuSampleVw} 
\end{align}
or some other  $t^{\mathrm{in}}$-dependent form.
When a single execution is split into two parts:  $dV=dV^{(a)}+dV^{(b)}$,
the measure (\ref{dVmuSample}) preserves the moments $\Braket{Q_j \frac{dV}{dt}}$,
whereas the original measure (\ref{dvpmuSample}) does not preserve $\Braket{Q_j \frac{d\varpi}{dt}}$.
In this sense, the modified measure (\ref{dvpmuSampleVw}) does preserve the moments.

When $t^{\mathrm{in}}$ is available, a potentially much better option than (\ref{dvpmuSampleVw}) is to calculate the same 
$dV/dt$, but assume the event occurred at the time of the initiating order arrival.
This would be the same expression (\ref{momentsDef}), with the same execution volumes $dV$,
but the specific $dV$'s are now re-ordered according to the initiating order arrival time.
Whereas in (\ref{dVmuSample}) the volumes $dV$
correspond to events that occurred at the moments of execution, the volumes $dV^{\mathrm{in}}$
now correspond to events that occurred at the moments of the initiating order arrival.
The measure
\begin{align}
d\eta = dV -dV^{\mathrm{in}}  \label{dvdvin} 
\end{align}
represents the in/out balance of LOB execution.
For a state with a large positive $d\eta$,
the trades are mostly created from long-standing LOB orders.
A state with a negative $d\eta$ corresponds to the arrival of orders to the LOB that will be executed later.
One may consider the actual volumes for initiating and matched orders,
but the LOB volume is easy to manipulate \cite{2016arXiv160305313G}.
Thus, it is expected to be a better choice to use the execution volume $dV$ and simply
re-order the execution events according to the moments of the initiating order arrival:
The measure $dV$ is a discrete measure corresponding to execution events that occurred at the actual execution time $t$,
whereas the measure $dV^{\mathrm{in}}$
corresponds to the same executions,
but as if they had occurred at the time of the initiating order arrival to the LOB, $t^{\mathrm{in}}<t$.
If $t^{\mathrm{in}}=t$, then such executions do not contribute to $d\eta$.
If a single large market order matches several limit orders in the LOB,
or several simultaneously arrived market orders match a single limit order in the LOB,
this creates several $dV$, $dV^{\mathrm{in}}$ contributions to the measure,
some of which may have identical $t$ or $t^{\mathrm{in}}$.
The preliminary analysis of the dynamics of $d\eta/dt$ is presented in Appendix \ref{EtaFlow}.
Overall, it is similar to that of $dV/dt$ (\ref{IclassicDef}) and does not seem to provide any critical advantage.

Note that $d\eta$ (\ref{dvdvin}) tracks LOB dynamics regardless of the initiating order type: buy or sell.
Market observations show that, since 2008-2010, the initiating order type,
often used by market practitioners, is typically useless for market dynamics studies \cite{2015arXiv151005510G, 2016arXiv160305313G}.
The buy/sell order volume imbalance (both in the LOB and in execution)
typically provides information that is inferior to price changes.

A fast, efficient, and numerically stable implementation of all these moment calculations in an arbitrary basis $Q_j$ is rather complex and has been discussed in \cite{2015arXiv151005510G,malyshkin2022market}.
The implementation is available from \cite{polynomialcode}; see the classes
\texttt{\seqsplit{com/polytechnik/trading/\{QVMDataM,QVMDataL,QVMDataP\}.java}}
and
\texttt{\seqsplit{com/polytechnik/freemoney/\{CommonlyUsedMomentsMonomials,CommonlyUsedMomentsLaguerre,CommonlyUsedMomentsLegendreShifted\}.java}} for an implementation.

An alternative, though not fully rigorous, method of calculation that allows the use of additional measures beyond those in (\ref{measuresList}) is the ``secondary sampling'' approach \cite{MalMuseScalp},
in which a calculated value at $t_l$
is treated as $f_l$ in (\ref{momentsDef}),
considering it as if it were a measured observation.
This enables the calculation of a new range of moments.
For example, in \cite{MalMuseScalp}, the maximal eigenvalue of an eigenproblem (\ref{GEV}) was used as an integration measure
\begin{align}
d\mu&=\lambda^{[\max]}_l-\lambda^{[\max]}_{l-1}
\label{muSecondarySampling}
\end{align}
in a way similar to (\ref{measuresList}).
This secondary sampling calculation method is beyond the scope of the current paper and will be discussed elsewhere. In this paper, the only moment calculation technique that cannot be reduced to direct sampling (\ref{momentsDef}) is the calculation of $\frac{dP}{dt}\frac{dV}{dt}$ moments, as described in Appendix \ref{QQMCalculation} below.

Note that all the measures in (\ref{momentsDef}) allow us to calculate moments only of the first derivative,
such as $I = dV/dt$, $dP/dt$, and so on. Moments of second derivatives, such as $d^2P/dt^2$, $\frac{dP}{dt}\frac{dV}{dt}$, or
$dI/dt = d^2V/dt^2$ (the latter two being particularly important for our future considerations),
cannot be obtained from direct sampling.
We will discuss approaches for their calculation below.
For now, we assume that all necessary first-order derivative moments are calculable and present a few examples of useful calculations with them,
followed by a generalization toward a possible solution of the strong form of the ultimate market dynamics problem.

\section{\label{FinData}Available financial data and time scales}
In this section, we discuss the available market data,
which can be regarded as a form of \textsl{experimental data} against which any theory should be tested.
We consider this topic important and therefore include a dedicated section on market data --- more precisely,
on available trade execution data as a form of measured experimental evidence.
We developed an efficient method for computing the moments from this data,
which arrive as a continuous stream of individual trades. Our theoretical framework is built upon these moments.

The transaction sequence data $(t_l, P_l, dV_l)$ is available across various markets and time scales ---
from high-frequency exchange trading in liquid markets operating at sub-microsecond intervals,
to fixed-income over-the-counter markets with time scales of hours or even days,
and to real estate markets where transactions may take months to complete.
Derivatives, commodities, and emerging markets also exhibit their own specific characteristics.
In our approach, we require a liquid market with a large number of transactions and active participants.
The data must be of high quality and easily accessible at low or no cost.
For these reasons, the U.S. equities market is the natural first choice for applying our theory.

End-of-day market close data is freely available from numerous sources,
such as \href{https://finance.yahoo.com/}{Yahoo Finance} and various data aggregators.
However, daily close data is insufficient for applying our theory.
The concept of execution flow maximization requires analysis at the level of individual transactions
as they occur in real time from market participants.
Moreover, the use of ``daily close'' data introduces an artificial time scale (one day),
which undermines the key strength of our approach ---
the automatic selection of the relevant time scale based on the maximization of the execution flow.

The NASDAQ ITCH feed\cite{itchfeed} provides LOB data and full lifecycle information for each order ---
from its ``add order'' event to ``cancel'' or ``execute''.
However, the daily traded volume on NASDAQ represents only a fraction of the total daily traded volume of the U.S.
equities market. Moreover, the primary value of this feed --- the limit order book information ---
has become much less significant. Since approximately 2008--2010,
exchange trading has become increasingly similar to dark pool trading.
The most typically observed LOB pattern is \cite{ArxivMalyshkinMuse} that an added order spends almost no time in the LOB.
It is executed almost immediately (possibly partially), with the unmatched portion being shortly canceled;
if no execution occurs, the entire limit order is typically canceled immediately.
Empirical observations show that over 90\% of orders that reach the best price level at some point
are eventually canceled\cite{nasdaqord,2015arXiv151005510G}.
The current exchange fee structure makes LOB cancellations very cheap,
creating a significant incentive for trading algorithms to submit orders for purposes other than actual execution.
Executed orders (trades) provide much more meaningful information,
since completing a round trip -- buying and then selling an actual asset --
is considerably more costly and risky than simply adding and canceling orders in the LOB.
This is why, in this paper, we study the execution flow dynamics and not the LOB dynamics.

Moreover, current U.S. regulations require that all actual trades be published through the Consolidated Tape System (CTS),
which includes execution transactions from all exchanges and dark pools.
Historical tapes, known as daily TAQ (Trade and Quote), can be acquired from NYSE\cite{NYSEtaq}
at a reasonable cost, or some free samples can be downloaded
from their website at \url{https://www.nyse.com/market-data/historical/daily-taq}.
A single daily TAQ file typically contains over 100 million execution transactions (lines)
and exceeds 10 GB in uncompressed size.
Across all tickers, the daily volume calculated from the daily TAQ files is slightly higher than
the value reported by Yahoo Finance and significantly larger than that computed from the NASDAQ ITCH daily file.

In this paper, we primarily use data from NYSE daily TAQ.
For the purpose of comparison with our previous works,
we also use data from Nasdaq ITCH for September 20, 2012.
This date was selected in \cite{2015arXiv151005510G} for a simple reason:
the market exhibited a bear trend before 10:00 and a bull trend with high volatility afterward.
Such market behavior often leads to significant losses for automated trading machines.

For the purpose of testing, this market data can be viewed as a large \texttt{.csv} file with lines of the form:
\begin{align}
\text{\noindent\parbox[t]{0.9\columnwidth}{\tt
|ticker\ \ \ \ \ \ time\ \ \ \ \ \ \ price\ \ \ \ shares \\
NVDA\ \ \ \ 31556271038450\ \ 156.26\ \ \ \ 3\\
TSLA\ \ \ \ 31556274115189\ \ 298.7\ \ \ \ \ 109\\
TQQQ\ \ \ \ 31556285245282\ \ 81.88\ \ \ \ \ 5\\
TQQQ\ \ \ \ 31556335367235\ \ 81.8899\ \ \ 5\\
PLTR\ \ \ \ 31556335813084\ \ 135.48\ \ \ \ 2\\
TSLA\ \ \ \ 31556519786918\ \ 298.675\ \ \ 1\\
NVDA\ \ \ \ 31556540197765\ \ 156.27\ \ \ \ 1\\
TSLA\ \ \ \ 31556542897531\ \ 298.6981\ \ 3\\
AAPL\ \ \ \ 31556561439699\ \ 207.2099\ \ 6\\
TSLA\ \ \ \ 31556591750551\ \ 298.7\ \ \ \ \ 20\\
TSLA\ \ \ \ 31556595205403\ \ 298.7\ \ \ \ \ 5\\
PLTR\ \ \ \ 31556602938660\ \ 135.48\ \ \ \ 5\\
TSLA\ \ \ \ 31556640789406\ \ 298.7\ \ \ \ \ 45
}}
\label{inpData4Columns}
\end{align}
Each line contains the ticker, execution time (in nanoseconds since midnight), execution price, and the number of shares traded.
Such a file can be readily computed from NASDAQ ITCH or NYSE daily TAQ, see Appendix \ref{SoftwareDescription} below.
The strong form of the ultimate market dynamics problem is equivalent to the existence of a stream processor
(possibly with an internal state) that reads such a file line-by-line, updates its internal state,
and posts trades that consistently result in a positive P\&L.
As emphasized earlier \cite{2015arXiv151005510G}, the price prediction problem is distinct from P\&L prediction;
we will discuss this difference below.
For now, let us note that all moments of the form (\ref{measuresList}) can be efficiently calculated from such a stream using an incremental recurrent update and a Newton-binomial type expansion:
\begin{align}
Q_j(ax+b)=\sum_{k=0}^{j}c_k Q_k(x) \label{NewtonBinomial}
\end{align}
This generalizes the familiar expression $(1+x)^j=\sum_{k=0}^{j}C_{j}^{k}x^k$ to an arbitrary polynomial basis $Q_j$.
See \cite{polynomialcode} for implementation details,
especially \texttt{\seqsplit{com/polytechnik/utils/RecurrenceAB.java:getBasisFunctionsOnPolynomialArgument}},
which implements, in a general orthogonal basis defined by the given three-term recurrence coefficients,
the expansion of a polynomial function of a polynomial into a single polynomial.
The exponential weight $\omega=\exp((t-t_{now})/\tau)$ makes the incremental calculation of moments straightforward;
if a fixed-window weight function were used, the recurrent calculations would become problematic.

\begin{figure}[t]
% java com/polytechnik/algorithms/TestCall_FreeMoneyForAll --musein_file=dataexamples/aapl_old.csv.gz --musein_cols=9:1:2:3  --n=12 --tau=128 --measure=CommonlyUsedMomentsLegendreShifted --museout_file=/u1/tmp//pdata/res/movingaverage/Aver128_museout.dat
%
% fn128="/u1/tmp//pdata/res/movingaverage/Aver128_museout.dat"; fn512="/u1/tmp//pdata/res/movingaverage/Aver512_museout.dat" ; set output "q.eps"  ;set pointsize 5 ;  set key bottom right ;set terminal postscript eps size 12cm,8cm enhanced color font 'Times,18'; set xrange [9.7:10.3]; set yrange [693:699] ; set grid front ; plot  fn128 using ($1/3600e9):($4) with lines lc 2 lw 3 title "{{/Times-Oblique P}^{{/Symbol-Oblique t}=128}" , fn128 using ($1/3600e9):($4-$5) with lines lt 1 lc 2 lw 0.5 notitle , fn128 using ($1/3600e9):($4+$5) with lines lt 1 lc 2 lw 0.5 notitle  , fn512 using ($1/3600e9):($4) with lines lt 1 lc rgb "#BBCD32" lw 3 title "{{/Times-Oblique P}^{{/Symbol-Oblique t}=512}" , fn512 using ($1/3600e9):($4-$5) with lines lt 1 lc rgb "#BBCD32"  lw 0.5 notitle , fn512 using ($1/3600e9):($4+$5) with lines lt 1 lc rgb "#BBCD32" lw 0.5 notitle , fn128 using ($1/3600e9):($3) with lines lt 1 lc 1 lw 3 title "{/Times-Oblique P}"
  \includegraphics[width=0.9\columnwidth]{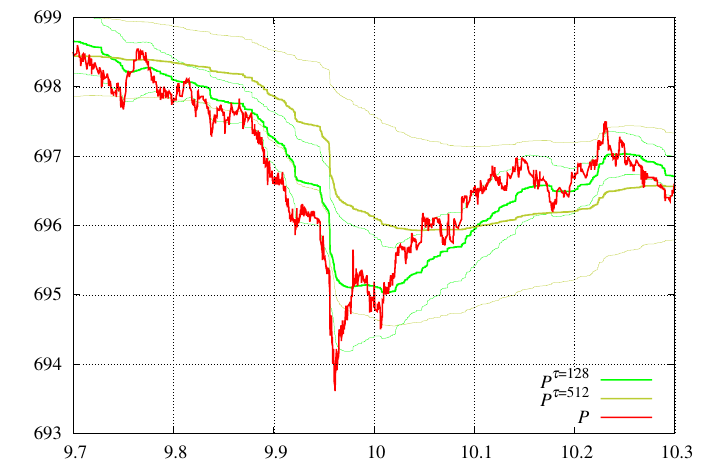}
  \caption{\label{MovingAveragePlot}
    An example of regular exponential moving average
    corresponding to $\tau=128$s and $\tau=512$s.
    Standard deviation is also calculated with the same $\tau$ and
    moving average $\pm$ standard deviation is plotted as a thin line in the same color.
    As $\tau$ increases -- the moving average ``shifts to the right''
    ($\tau$-proportional time delay, lagging indicator).
 The data is for AAPL stock on September, 20, 2012.
  }
\end{figure}

Let us provide a simple demonstration. Assume we have obtained three moments:
$\Braket{Q_0I}$, $\Braket{PQ_0I}$, and $\Braket{P^2Q_0I}$.
Since $Q_0$ is constant, these correspond (up to a constant factor)
to volume-weighted ($I = dV/dt$) entities $P^0$, $P^1$, and $P^2$, respectively.
The moments $\Braket{Q_0}$, $\Braket{PQ_0}$, and $\Braket{P^2Q_0}$
represent time-weighted $P^0$, $P^1$, and $P^2$.
Using any of these moments, one can construct a moving average (\ref{pMovaverage})
and a moving standard deviation (\ref{sigmaMovaverage}).
In Fig. \ref{MovingAveragePlot}, two volume-weighted moving averages are calculated for $\tau = 128$s and $\tau = 512$s.
The time-weighted moving average would be slightly smoother than the volume-weighted version.
The $x$-coordinate, consistent with our previous works, is expressed as a decimal fraction of an hour;
for example, $9.75$ in plot corresponds to 9:45 am. A $\pm$ single moving standard deviation is also shown in the plot.
As expected, the moving average is delayed (shifted to the right) by a time scale proportional to $\tau$
relative to the actual price, making it a lagging indicator.
When the input data undergoes a qualitative regime change, it takes a $\tau$-proportional lag for the moving average to reflect this transition.
Some popular trading strategies use events when the price crosses its moving average as triggers for action.
In \cite{2015arXiv151005510G}, we discuss the shortcomings of such approaches when operating on a single time scale.

As a demonstration, let us present another perspective on the meaning of the moving average.
Consider not $3$, but $2n+1$ moments $\Braket{P^k  I}$, with $k=0 \dots 2n$.
Now consider the problem of constructing a polynomial of degree $n$
that satisfies the optimization problem of minimizing the square
of the polynomial with respect to the $L^2$ measure relatively $\Braket{\cdot}$:
\begin{align}
\Braket{\left(P^n+a_{n-1}P^{n-1}+a_{n-2}P^{n-2}+\dots+a_0\right)^2  I}\to\min
\label{OpolMin}
\end{align}
The solution yields an orthogonal polynomial of degree $n$ constructed with respect to the given measure.
The roots $\pi^{[i]}$ of this polynomial can be found by solving the following generalized eigenproblem:
\begin{align}
\sum\limits_{k=0}^{n-1}\Braket{P^j|PI| P^k} \alpha_k^{[i]} &=
\pi^{[i]} \sum\limits_{k=0}^{n-1}\Braket{ P^j |I|P^k} \alpha_k^{[i]}
\label{OpolMinRoots}
\end{align}
Here we have changed the notation to Paul Dirac \href{https://en.wikipedia.org/wiki/Bra%E2%80%93ket_notation}{bra--ket notation},
a form that will be very useful below. For real matrices, we simply have
$\Braket{P^j|PI| P^k}=\Braket{P^{j+k+1}I}$, and $\Braket{P^j|I| P^k}=\Braket{P^{j+k}I}$.
As long as the right-hand side matrix $\Braket{P^j|I| P^k}$ is positively definite, the problem has $n$ solutions.
The $n$ eigenvalues $\pi^{[i]}$ of the eigenproblem (\ref{OpolMinRoots})
correspond to the $n$ roots of the degree-$n$ polynomial defined in (\ref{OpolMin}).
The roots $\pi^{[i]}$ correspond to the Gaussian quadrature nodes
that interpolate the measure used to construct the polynomial with an $n$-point discrete measure.
The corresponding weights $w^{[i]}$ can be obtained from the eigenvectors $\alpha^{[i]}$
by evaluating them at corresponding $\pi^{[i]}$;
alternatively, they can be determined from the Christoffel function. The sum of all weights $w^{[i]}$ equals $\Braket{I}$.
This is a common method for constructing
orthogonal polynomials from a given measure\cite{totik} and for finding their roots along with the corresponding measure weights.

\begin{figure}[t]
  \includegraphics[width=0.9\columnwidth]{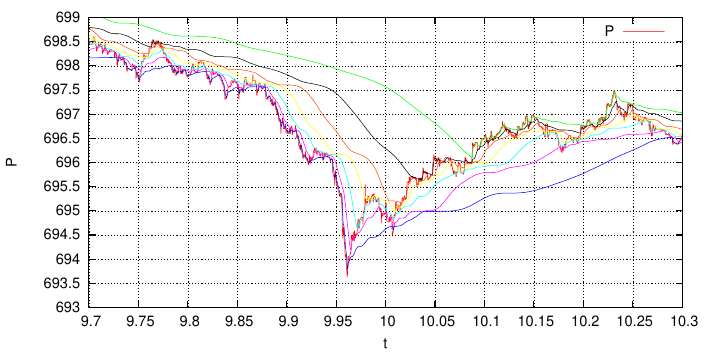}
  \caption{\label{opolMulti}
An example of a higher-order orthogonal polynomial root calculated from the moments
$\Braket{P^k  I}$, $k = 0 \dots 2n$, is shown for $n = 7$ and $\tau=128$s.
Seven roots are obtained, with a substantial volume expected to be traded at each corresponding price level.
In this example, the actual measure is approximated by a discrete measure with $n = 7$ support points.
The figure is reproduced from \cite{2015arXiv151005510G}.
  }
\end{figure}

One can note that the moving average (\ref{pMovaverage}) corresponds to the root of an orthogonal polynomial of degree $n = 1$, which has a single root;
the corresponding weight for this node is $\Braket{I}$.
Given a sufficient number of moments $\Braket{P^k  I}$,
which can be calculated from the market data as above,
one can construct higher-order polynomials and determine their roots.
A demonstration from Ref. \cite{2015arXiv151005510G} is shown in Fig. \ref{opolMulti} of this paper for $n = 7$ roots of a polynomial calculated from the moments $\Braket{P^k  I}$, $k = 0 \dots 2n$ with $\tau=128\mathrm{s}$.
These roots serve as the nodes of a Gaussian quadrature,
which approximates the measure used to construct the orthogonal polynomial with a discrete measure at $n$ support points.
A quadrature with $n = 1$ corresponds to a moving average,
while a quadrature with $n = 2$ (two nodes) provides not only the average
but also allows the estimation of the distribution’s median and skewness.
This is an example of constructing orthogonal polynomials for a single asset price.

For multiple assets (assuming the price phase space is relatively stable),
such an approach is not directly applicable.
A possible alternative is to construct the Christoffel function in the price space of several assets,
in a manner similar to that described in Appendix \ref{ChristoffelSpectrum}.
For a full basis, this approach provides an analogue of the joint price distribution;
selecting a few states with large coverage could potentially create a predictive model (provided the distribution is stable).
However, this approach -- similar to an orthogonal polynomial model -- is not dynamic;
it is more akin to returning to frequently visited points in the phase space.

These demonstrations are simple examples illustrating the potential use of a large number of moments.
While they operate on prices and generate charts, they do not directly convey information about market dynamics.
Nevertheless, the availability of a large number of sampled moments is valuable,
as it allows us to formulate and solve generalized eigenproblems, such as (\ref{OpolMinRoots}).
This specific eigenproblem primarily serves to plot informative charts
that highlight the price levels at which substantial trading occurred in the past.

A useful application of this orthogonal polynomial technique for market practitioners is as follows.
Instead of relying on the commonly used symmetric plots $P_{ma} \pm \sigma_{ma}$ to determine action thresholds,
a substantially better approach is to construct an orthogonal polynomial of degree $2$ or $3$
and monitor the crossing of the current last price with the minimum or maximum roots $\pi^{[i]}$ of the
corresponding orthogonal polynomial.
These roots correspond to the support points of trading volume and can capture distribution asymmetry
and other relevant factors, providing a more informative basis for trading decisions.
However, our aim is far more ambitious --- understanding market dynamics ---
and this example was presented solely to illustrate the eigenproblem technique that we actively employ in the subsequent analysis.

\section{\label{ExecutionFlowCalculation}Execution Flow: Calculation and Methodology}
Execution flow $I=dV/dt$, the number of shares traded per unit time,
is a positive quantity --- a ratio of two measures $\omega dV$ and $\omega dt$ — and can be considered
as their Radon–Nikodym derivative. To calculate its value at a specific
point $x$, a number of approaches can be applied, from direct interval sampling
to a ratio of localized states\cite{2015arXiv151005510G}. Formally, even a least-squares
approach can be applied to interpolate $dV/dt$, for example,
\begin{align}
&\hspace{-3em} \Braket{\left(I-\sum_{j=0}^{n-1}\beta_j Q_j(x)\right)^2}\to \min \label{LSIproblem} \\
I_{LS}(x)&=\sum\limits_{j,k=0}^{n-1} Q_j(x) G^{-1}_{jk}\Braket{Q_kI} \label{Ils} \\
G_{jk}&=\Braket{Q_j |Q_k} \label{GramM}
\end{align}
where $G^{-1}$ is the inverse of the Gram matrix (\ref{GramM}).
This expansion uses $n$ moments $\Braket{Q_k I}$ and $2n-1$ moments $\Braket{Q_k}$ to compute.
This approach does not preserve the internal structure of the execution flow (for example, its inherently positive sign)
and does not incorporate the full past history in a way that allows determining thresholds,
such as whether the execution flow at $t_{now}$ is small or large.
Moreover, expanding the highly fluctuating $dV/dt$, which varies by many orders of magnitude,
in a polynomial basis discards the critical information contained in its spikes.

We need a general method to account for highly fluctuating values over the polynomial moments.
The idea is to interpolate not the observed value $I$, but the probability density.
Consider a function $\psi(x)=\sum_{j=0}^{n-1} \alpha_j Q_j(x)$ that defines the density $\psi^2(x(t))\omega(t)dt$,
and a value expressed as a ratio of two measures, such as $I = dV/dt$.
The value of $I$ corresponding to a given state $\psi(x)$ can then be estimated as measures ratio
\begin{align}
I_{\psi}&=\frac{\Braket{\psi|I|\psi}}{\Braket{\psi|\psi}}=
\frac{\sum\limits_{j,k=0}^{n-1}\alpha_j\Braket{Q_j|I| Q_k} \alpha_k}
{\sum\limits_{j,k=0}^{n-1}\alpha_j\Braket{Q_j| Q_k} \alpha_k} \label{Ipsi}
\end{align}
Here, we continue to use bra–ket notation;
for real matrices, we have
$\Braket{Q_j|I| Q_k}=\Braket{Q_jQ_kI}$, and $\Braket{\psi|I|\psi}=\Braket{\psi^2I}$.
The (\ref{Ipsi}) expansion uses $2n-1$ moments $\Braket{Q_k I}$ in the numerator and $2n-1$ moments $\Braket{Q_k}$ in the denominator.
The Gram matrix $\Braket{Q_j|Q_k}$ is obtained from $\Braket{Q_k}$ using the multiplication operator $c_m^{jk}$.
\begin{align}
  Q_j Q_k&=\sum_{m=0}^{j+k}c_m^{jk}Q_m
  \label{multiplicationOperator}
\end{align}
Its form is straightforward for monomial and Chebyshev bases, but can be quite challenging in other cases.
See our previous works and the code in \cite{polynomialcode} for implementation details,
especially \texttt{\seqsplit{com/polytechnik/utils/RecurrenceAB.java:getBasisFunctionsMultipliedByPolynomial}},
which implements the multiplication of a general orthogonal polynomial basis defined
by the given three-term recurrence coefficients.
Below, we will assume that any matrix $\Braket{Q_j|f| Q_k}$ for $j,k = 0 \dots n-1$
can always be obtained from the moments $\Braket{Q_m f}$, $m=0\dots 2n-2$ with (\ref{multiplicationOperator}).

In \cite{malyshkin2019radonnikodym}, we considered various forms of $\psi(x)$ to interpolate some value in two stages:
first, obtaining a wavefunction state satisfying certain requirements (such as a state $\psi_y(x)$ localized at $x = y$),
and then computing the Radon-Nikodym derivative in that state.
We do not require this interpolation theory here.
The only important feature of (\ref{Ipsi}) in the present context is that it is
a ratio of two quadratic forms of equal dimension $n$, i.e., it is a Rayleigh quotient.
If at least one of the two matrices, $\Braket{Q_j|I| Q_k}$ or $\Braket{Q_j| Q_k}$ in (\ref{Ipsi}),
is positively definite, then they can be simultaneously diagonalized via a generalized eigenproblem.
\begin{align}
\left|I\middle|\psi^{[i]}\right>&=\lambda^{[i]}\left|G\middle|\psi^{[i]}\right> \label{GEVd} \\
\sum\limits_{k=0}^{n-1}\Braket{Q_j|I| Q_k} \alpha^{[i]}_k&=
\lambda^{[i]}\sum\limits_{k=0}^{n-1}\Braket{Q_j| Q_k} \alpha^{[i]}_k \label{GEV} \\
\psi^{[i]}&=\sum\limits_{k=0}^{n-1}\alpha^{[i]}_k Q_k \label{psiEVI}
\end{align}
Eq. (\ref{GEVd}) is the bra--ket form of the explicit matrix form (\ref{GEV}).
This eigenproblem provides a solution for determining whether the current execution flow $I$ is low or high:
one can simply compare it's magnitude with the eigenvalues $\lambda^{[i]}$, e.g., if the value is close to the $\lambda^{[\mathrm{maxI}]}$,
the current $I$ is very high. In most situations, we are interested in determining whether the execution flow ``now'',
in the state $\psi_0$, is low or high.
In this case, it is often more convenient to consider the state projection
$\Braket{\psi_0| \psi^{[\mathrm{maxI}]}}^2$, where
\begin{align}
\psi_0(x)&=const \cdot\sum_{j,k=0}^{n-1}Q_j(x_0)G^{-1}_{jk}Q_k(x)
\label{psi0def}
\end{align}
is the state localized at $x_0$ corresponding to $t_{now}$,
rather than comparing $I_0=\Braket{\psi_0|I|\psi_0}$ with $\lambda^{[\mathrm{maxI}]}$.
However, this is an implementation detail, and the most important features of (\ref{GEV}) are:
\begin{itemize}
\item Given a sufficiently large $n$, it contains information about long-past $I$ values.
The eigenproblem matrices in (\ref{GEV}) incorporate different time scales,
with the range of ``stored'' time scales determined by the value of $\tau$ and the problem dimension $n$.
The corresponding realization of an observable in the state $\psi(x)$ is given by the Rayleigh quotient (\ref{Ipsi}).
\item The measures $\omega dV$ and $\omega dt$ enter symmetrically; there are two matrices forming the Rayleigh quotient.
To compute the left- and right-hand side matrices in eigenproblem (\ref{GEV}), $2n-1$
moments $\Braket{Q_j I}$ and $\Braket{Q_j}$ are required for each matrix respectively.
\item The problem inherently contains thresholds (the eigenvalues $\lambda^{[i]}$),
making it particularly simple to determine whether the current value is low or high.
\item For large enough $n$, the method can handle large spikes.
The approach separates probabilities and values: the situation is analogous to quantum mechanics,
where a single ``several-orders-off'' state essentially does not affect the result if its probability is near zero.
This contrasts with $L^2$ approaches, such as in (\ref{LSIproblem}), where a single ``several-orders-off''
observation can completely distort the result.
\item The eigenvectors (\ref{psiEVI}) have algebraic properties that are important for our subsequent considerations.
\end{itemize}

The approach described is a very general method that can be applied to any observable representable
as a Radon–Nikodym derivative $d\mu/d\nu$.
One simply constructs two matrices,  $\Braket{Q_j|d\mu/dt|Q_k}$ and $\Braket{Q_j|d\nu/dt|Q_k}$,
corresponding to the numerator and denominator measures, and then solves the generalized eigenproblem (\ref{GEV}).
See Ref. \cite{malyshkin2022market}, Section III, which presents a table of different left- and right-hand
side matrices we previously considered.
As discussed in \cite{2015arXiv151005510G}, when applied to market dynamics,
the execution flow $I = dV/dt$ -- a highly fluctuating quantity -- is the most important characteristic;
it corresponds to $d\mu=dV$ and $d\nu=dt$.
The left- and right-hand matrices in (\ref{GEV}) are obtained as
$Q_jQ_k$
averaged with the  $\omega d\mu$ and $\omega d\nu$ measures.
When $Q_j=Q_j(x(t))$ and $d\nu=dt$, the analytic value for the right-hand side matrix in (\ref{GEV}) can be obtained;
this both simplifies calculations and improves numerical stability.

If full information of the LOB is available, then, instead of the $\omega dt$
averaging for the right-hand side, one may use averaging with some other measure.
Each execution is now characterized by the time $t$, price $P$, matched volume $dV$, and the time $d\varpi=t_l-t_l^{\mathrm{in}}$
the initial limit order spent in the LOB before being matched --
with this data available, the ultimate market dynamics problem has a sequence of quadruples as input.
\begin{align}
&(t_l, P_l, dV_l, t_l^{\mathrm{in}}) & l=1\dots M
\label{Quadruples}
\end{align}
Specifically, in (\ref{GEV}),
the matrices are now averaged on the same grid (all execution events):
the left-hand side matrix with the measure
$\omega dV$,  and the right-hand side matrix uses some other measure, see Appendix \ref{DirAssym} below.
This allows for the inclusion of additional market timing information
that may not already be incorporated into the current price \cite{2016arXiv160305313G}.
However, whereas the execution volume $dV$
is required to be provided by exchanges and dark pools \cite{NYSEtaq},
there is no such requirement for $t_l^{\mathrm{in}}$.
One can either use
a proxy, or obtain $d\varpi$
for each execution directly from LOB events,
as in \cite{itchfeed}, with the exception of ``hidden'' trades.
However, the LOB data from exchanges typically has far fewer execution transactions than the full TAQ\cite{NYSEtaq}.

Note that the eigenproblem (\ref{OpolMinRoots}) considered earlier has a similar structure to (\ref{GEV}),
but is applied to price $P$,
with the matrices $\Braket{Q_j|PI|Q_k}$ and $\Braket{Q_j|I|Q_k}$. The resulting eigenvalues indicate price levels with high traded volume. For a general basis $Q_j$, this will no longer correspond to an orthogonal polynomial; however,
by setting $Q_j(x(t)) = P^j(t)$ and $d\mu = PdV$ and $d\nu=dV$,
one recovers (\ref{OpolMinRoots}) exactly from (\ref{GEV}).
With $d\mu = Pdt$ and $d\nu=dt$, one also recovers (\ref{OpolMinRoots}),
but the eigenvalues now indicate the price levels at which the most time was spent.\footnote{
\label{measureFromMinLambda}
The measures $dV$ and  $dt$ are positive and allow for obtaining a large number of eigenvectors.
We want to note that if we consider a non-positive measure $d\mu=dV-\lambda^{[\min]}dt$,
then with this measure, only a limited number of orthogonal polynomials can be constructed.
For example, if the basis is $Q_j(x(t)) = P^j(t)$ and $\lambda^{[\min]}$
is the minimal eigenvalue of (\ref{GEV}) in this basis,
then for the measure $d\mu=dV-\lambda^{[\min]}dt$, the $n$-th orthogonal polynomial has zero $L^2$ norm in (\ref{OpolMin}),
(where  $dV/dt$ should be replaced by $d\mu/dt$) and coincides with the eigenvector of the eigenproblem (\ref{GEV})
corresponding to $\lambda^{[\min]}$.
See Ref. \cite{2015arXiv151005510G}, where the measure $d\mu=dV-\lambda^{[\min]}dt$
is considered in different bases for various applications.
For example, it immediately follows that the eigenvector corresponding to the minimal/maximal eigenvalue
of eigenproblem (\ref{GEV}) of dimension $n$ has exactly $n-1$ roots, but not necessarily on the support of the measures  $dV$ or $dt$.
The orthogonal polynomials constructed on a positive measure have all their roots on the support of the measure.
}

Now we present several simple demonstrations of execution flow properties computed from exchange data.
Our goal is to illustrate the approach in a way similar to the industry-standard ``moving average'' concept.
We use the basis
$x=\exp((t-t_{now})/\tau)$, $\omega=\exp((t-t_{now})/\tau)$, and $Q_j(x)$ as a polynomial of degree $j$
(the result is invariant with respect to the specific choice of polynomial basis).
Using these data, we compute $2n - 1$ moments $\Braket{Q_m I}$ by direct sampling (\ref{momentsDef}).
The calculations are performed at each time $t$ over the interval preceding current $t_{now}$ -- analogous to a moving average --
with $t_{now}$ advancing through the sample.
The moments $\Braket{Q_m}$ are known analytically for the chosen $x$ and $\omega$.
All these moments are then used to formulate the eigenproblem (\ref{GEV}) and obtain the eigenvalues $\lambda^{[i]}$
and eigenvectors $\psi^{[i]}(x)$.
Finally, we compute the price $P$ and $t-t_{now}$ in the state $\psi^{[\mathrm{maxI}]}$ corresponding to the maximal eigenvalue
$\lambda^{[\mathrm{maxI}]} = \Braket{\psi^{[\mathrm{maxI}]}|I|\psi^{[\mathrm{maxI}]}}$, the states are normalized
as $\Braket{\psi|\psi} = 1$.
\begin{align}
P^{[\mathrm{maxI}]}&=\frac{\Braket{\psi^{[\mathrm{maxI}]}|PI|\psi^{[\mathrm{maxI}]}}}{\Braket{\psi^{[\mathrm{maxI}]}| I|\psi^{[\mathrm{maxI}]}}} \label{PIH} \\
T^{[\mathrm{maxI}]}&=\frac{\Braket{\psi^{[\mathrm{maxI}]}|\frac{t-t_{now}}{\tau}I|\psi^{[\mathrm{maxI}]}}}{\Braket{\psi^{[\mathrm{maxI}]}|I|\psi^{[\mathrm{maxI}]}}} \label{TIH}
\end{align}
The value of $P$ in the $\psi^{[\mathrm{maxI}]}$ state (\ref{PIH}) is an important characteristic
of our approach to market dynamics\cite{2015arXiv151005510G}.
The $t-t_{now}$ in this state (\ref{TIH}) has a much simpler structure than $P$ and allows
a straightforward visualization of qualitative ``switching'' in the structure of the $\psi^{[\mathrm{maxI}]}$ state.
While the moments $\Braket{Q_mI}$ and $\Braket{Q_mPI}$ are just glorified moving averages,
the $P^{[\mathrm{maxI}]}$ and $T^{[\mathrm{maxI}]}$ are not. There is an additional step --
selecting the state $\psi^{[\mathrm{maxI}]}$ from the (\ref{GEV}) solutions.
Thus, the $P^{[\mathrm{maxI}]}$ (or $T^{[\mathrm{maxI}]}$) can be viewed as a moving average with internal degrees of freedom.

A regular moving average is computed on a past sample by averaging an observable
with a density such as $\omega(t)dt$, which remains the same.
A moving average with internal degrees of freedom is computed on
a past sample by averaging an observable with a density such as $\psi^2(x(t))\omega(t)dt$,
which changes (according to some equation) as new observations are processed.
This is similar to the two-stage Radon-Nikodym approach of Ref. \cite{malyshkin2019radonnikodym}:
first select the state, and then evaluate the observable in that state.
For market dynamics, the $\psi(x)$ in the integration density
is governed by the generalized eigenproblem (\ref{GEV}); the $\psi(x)$ in question is its maximal eigenvector.

\begin{figure}[t]

\includegraphics[width=0.9\columnwidth]{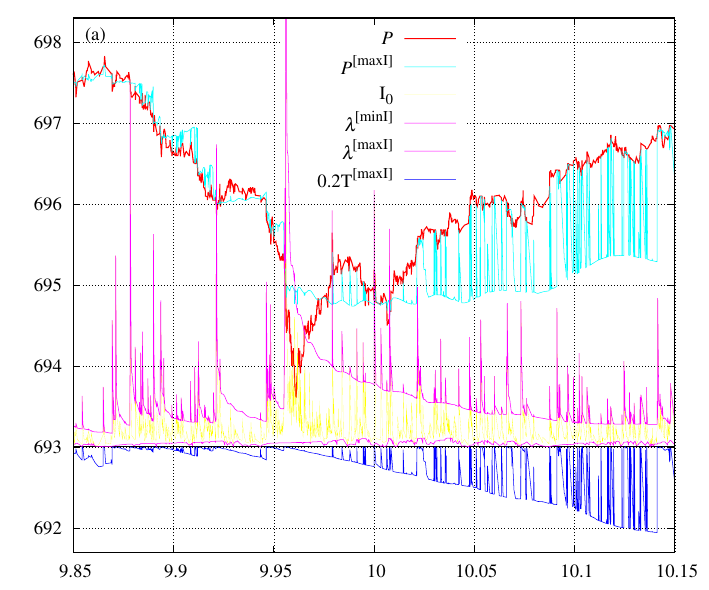}
\includegraphics[width=0.9\columnwidth]{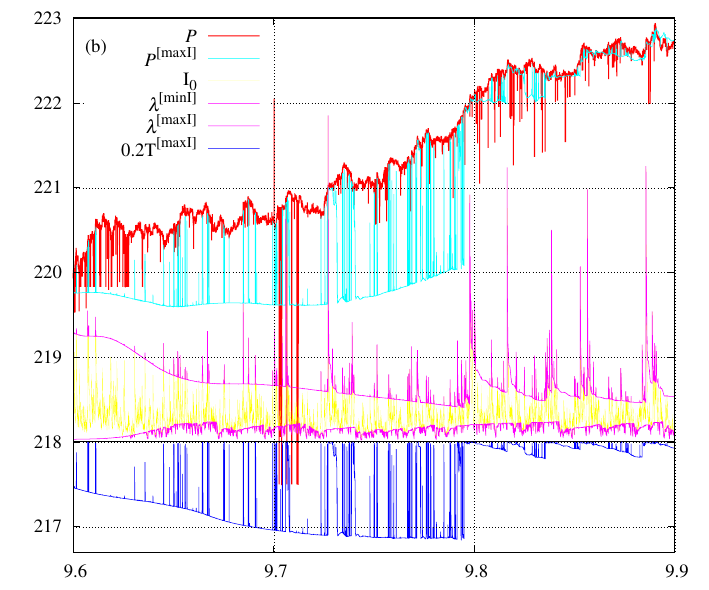}
  \caption{\label{IHEGCalc}
A demonstration of execution flow.
(a) AAPL, September 20, 2012. (b) AAPL, April 1, 2025.
We present the original price $P$ and $P^{[\mathrm{maxI}]}$ (\ref{PIH}) (light blue).
The other plots are shifted upward (to 693 for panel (a) and 218 for panel (b)),
and the execution flow is scaled (by $5\cdot10^{-6}$ for (a) and $10^{-6}$ for (b)) to avoid cluttering the chart.
We also present $T^{[\mathrm{maxI}]}$ (\ref{TIH}), the minimal and maximal eigenvalues of (\ref{GEV}),
and $I_0 = \Braket{\psi_0|I|\psi_0}$ (yellow); the result is obtained for $n=12$ and $\tau=128$s.
One can clearly observe an immediate switch due to internal degrees of freedom in (\ref{GEV}),
without the $\tau$-proportional lag typical of regular moving averages shown in Fig. \ref{MovingAveragePlot}.
The most important observation in these plots is the occurrence of price singularities at time moments when the execution flow is large, 
i.e., when $I_0$ is close to $\lambda^{[maxI]}$.
  }
\end{figure}

In Fig. \ref{IHEGCalc}, for the same AAPL dataset as in the figures above,
we present $P^{[\mathrm{maxI}]}$ and $T^{[\mathrm{maxI}]}$, along with the maximal and minimal eigenvalues of (\ref{GEV}).
The value of $I$ at $t_{now}$, evaluated as $I_0=\Braket{\psi_0|I|\psi_0}$, is also shown.
Note that $P^{[\mathrm{maxI}]}$, $T^{[\mathrm{maxI}]}$, $\lambda^{[\mathrm{minI}]}$, and $\lambda^{[\mathrm{maxI}]}$ are moving averages with internal degrees of freedom:
the state is determined by the eigenvalue problem (\ref{GEV}).
The $I_0 = \Braket{\psi_0|I|\psi_0}$
can be regarded as a traditional moving average (with a very small effective scale of about $\tau/(2n)$),
since $\psi_0(x)$ (\ref{psi0def}) does not change with the data.
Contrary to a regular moving average, where it takes a $\tau$-proportional lag to reflect a qualitative regime change
(see Fig. \ref{MovingAveragePlot}), a moving average with internal degrees of freedom exhibits an immediate ``switch''.
It is convenient to look at $T^{[\mathrm{maxI}]}$ (\ref{TIH}), which grows almost linearly when there is no spike in execution
flow and drops to nearly zero during an execution spike, when $\psi^{[\mathrm{maxI}]}(x(t))$ is localized near $t_{now}$
(i.e. when $\Braket{\psi^{[\mathrm{maxI}]}|\psi_0}^2=\left|\frac{\psi^{[\mathrm{maxI}]}(x_0)}{\psi_0(x_0)}\right|^2$ is close to $1$).
The most important observation in these plots is the occurrence of price singularities at time moments when the execution flow is large, 
i.e., when $I_0$ is close to $\lambda^{[maxI]}$.
It is this empirically observed fact that allows us to state that the execution flow, $I=dV/dt$,
not the traded volume $V$, is the driving force of the market. This is the main ideological distinction from the commonly studied market impact concept (price sensitivity to traded volume) \cite{bucci2019crossover}.

The equation (\ref{GEV}) for $\max I$, along with $P^{[\mathrm{maxI}]}$,
is the result we obtained back in \cite{2015arXiv151005510G}.
We even constructed a trading strategy that prevents catastrophic losses. The key idea is to predict $I=dV/dt$, not price.
This approach is very accurate: if there is a liquidity excess event
(current $I_0$ is large, i.e., $\Braket{\psi^{[\mathrm{maxI}]}|\psi_0}^2>0.9$), then future $I_0$ will be low.
Similarly, if there is a liquidity deficit event
(current $I_0$ is low, i.e., $\Braket{\psi^{[\mathrm{minI}]}|\psi_0}^2>0.9$), then future $I_0$ will be high.
This may seem trivial -- alternating periods of low and high liquidity -- but it demonstrates that liquidity (not price) undergoes large oscillations
(spanning several orders of magnitude),
with price changes (typically up to about one percent) 
being a consequence of these liquidity fluctuations.
The key element of the strategy is that it trades liquidity:
providing liquidity during deficits and taking it during excesses.
Specifically, the trader should open a position during liquidity deficits and close it during liquidity excesses.
The rationale is simple: holding a zero position during liquidity excess makes the system resilient to adverse market moves,
while entering a position during liquidity deficits (when volatility is small) allows the strategy
to capture the majority of market movement.
Our experiments (both paper trading and actual NASDAQ trading in 2010–2012)
confirm that this is the only strategy we found that avoids eventual catastrophic P\&L loss.
A directional trading strategy that is not predisposed to catastrophic P\&L loss
must include \textsl{at least four types of events}:
\begin{itemize}
\item Open long position
\item Close existing long position
\item Open short position
\item Close existing short position
\end{itemize}
These events are ``actionable triggers''; they cause the trader (or an automated trading machine) to perform an action.
Note that a strategy with only two types of events (e.g., when ``close existing long'' is the same as ``open short'')
will inevitably fail eventually, resulting in catastrophic P\&L loss.
Equation (\ref{GEV}) indicates when to open a position (current $I_0$ is low) and when to close it (current $I_0$ is large).
As shown above, these conditions translate into projections of $\psi_0$ onto $\psi^{[\mathrm{maxI}]}$ and $\psi^{[\mathrm{minI}]}$.
Specifically: open a position at a low $I$, e.g. $\Braket{\psi^{[\mathrm{minI}]}|\psi_0}^2>0.9$,
and close a position at a high $I$, e.g. $\Braket{\psi^{[\mathrm{maxI}]}|\psi_0}^2>0.9$.
However, this strategy does not specify the direction of the position when opening: whether to go long or short?
One could potentially express this execution flow prediction through volatility trading with options,
but this market is much less liquid, and transaction fees prevented us from performing experiments.

Since \cite{2015arXiv151005510G}, we have devoted substantial effort to determining the direction:
whether to open long or short when $I_0$ is low?
The best directional indicator we found back then, and failed to improve in subsequent works,
is the difference between the last price $P^{last}$ and $P^{[\mathrm{maxI}]}$ from (\ref{PIH}):
\begin{align}
\mathrm{dir}_{dPI}&=\lambda^{[\mathrm{maxI}]}\left(P^{last}-P^{[\mathrm{maxI}]}\right) \label{dirOld}
\end{align}
Check Fig. \ref{IHEGCalc}: you can see fast regime switches and effective tracking of execution flow.
Selecting a state other than $\psi^{[\mathrm{maxI}]}$  worsens the outcome; see Appendix \ref{StateSelectionProblem} below.

\begin{figure*}[t]
% /home/dev/scala/scala-2.13.18/bin/scala   com.polytechnik.algorithms.ExampleNoSupplyDemand dataexamples/aapl_old.csv.gz

%set output "q.eps" ; set pointsize 5 ; set terminal postscript eps size 12cm,6cm enhanced color font 'Times,22'; set label "(a)" at screen 0.17,0.90; set xrange [9.5:16]; set yrange [693:701] ; set xtics 0.5 ; set grid ;  set ytics 1 ; set ylabel "P" ;  set xlabel "t" ; l=694; s=0.5;cx=0.7; oo=0.7;ox=0.5; plot   "/u1/tmp//pdata/res/movingaverage/Aver128_museout.dat" using ($1/3600e9):($3) with lines lw 2 title "{/Times-Oblique P}(t)"
\includegraphics[width=0.95\columnwidth]{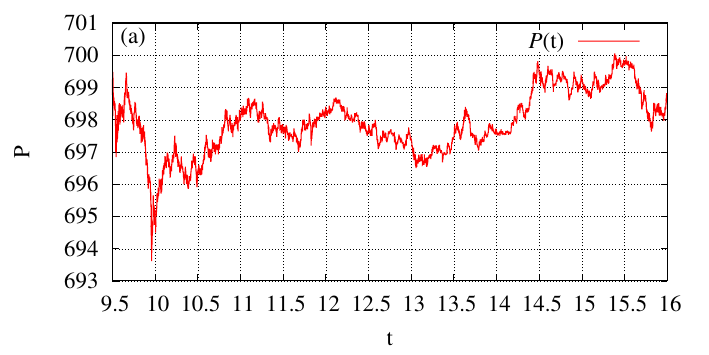}
%set output "q.eps" ; set terminal postscript eps size 12cm,6cm enhanced color font 'Times,22'; l=696; s=0.5;cx=0.7; oo=0.7;ox=0.5; set xrange [693:701] ; set xlabel "P" ;  set ytics 100000 ; set label "(b)" at screen 0.20,0.90; plot   "histogram.csv" using (($1+$2)/2):($3)  with boxes lw 3 title "Volume"
 \includegraphics[width=0.95\columnwidth]{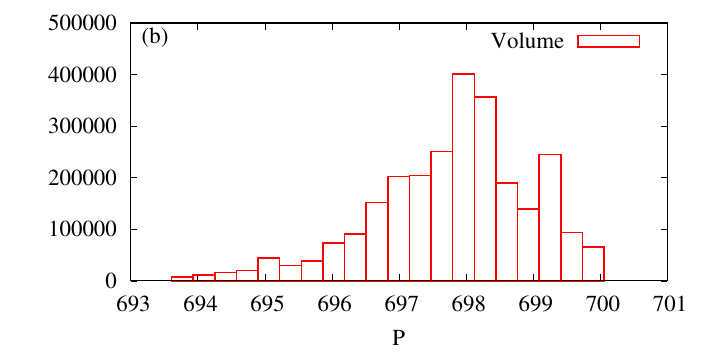} \\
%set output "q.eps" ; set terminal postscript eps size 12cm,6cm enhanced color font 'Times,22'; l=696; s=0.5;cx=0.7; oo=0.7;ox=0.5; set xrange [693:701] ; set grid ;set xlabel "P" ; set label "(c)" at screen 0.14,0.90; plot  "psi.csv" using ($1):($2)  with lines lc 3 lw 3  title "{/Times-Oblique I}_0({/Times-Oblique P})"
  \includegraphics[width=0.95\columnwidth]{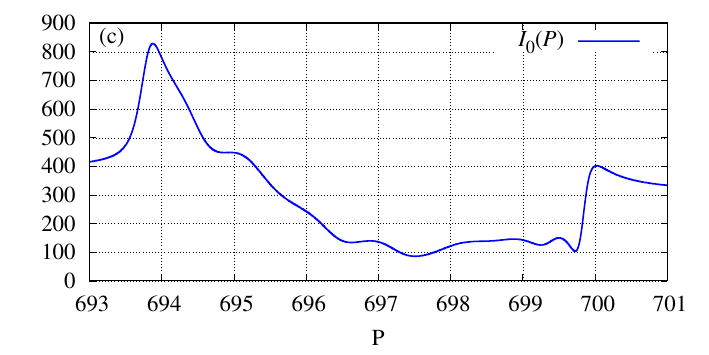}
%set output "q.eps" ; set terminal postscript eps size 12cm,6cm enhanced color font 'Times,22'; l=696; s=0.5;cx=0.7; oo=0.7;ox=0.5; set yrange [0:1] ; set xrange [693:701] ; set pointsize 4; set key spacing 2; set label "(d)" at screen 0.14,0.90; set grid ;set xlabel "P" ; plot  "psi.csv" using ($1):($6)  with lines lc 1 lw 6 title "<{/Symbol-Oblique y}^{[maxI]}|{/Symbol-Oblique y}_0>^2" ,   "psi.csv" using ($1):($4)  with lines lc rgb "#BBCD32" lw 6 title "<{/Symbol-Oblique y}^{[minI]}|{/Symbol-Oblique y}_0>^2"
   \includegraphics[width=0.95\columnwidth]{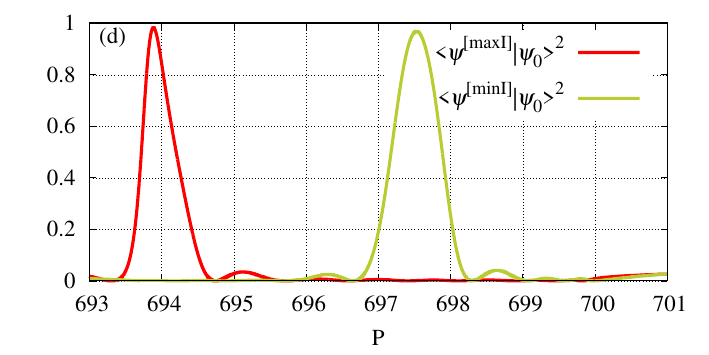}
\caption{\label{ntPlotChart}
AAPL stock for the same trading date, September 20, 2012, as in Fig. \ref{IHEGCalc}a.
(a): AAPL price for the entire trading day.
(b): Price histogram for volume distribution; there is a maximum around $P=698$ (about median); there is no price singularity around this level.
(c): $I_0(P)=\Braket{\psi_0|I|\psi_0}$ (shares/sec)
as a function of price; the singularities are clearly observed at the 
$694$ and 
$700$ price levels, which match the singularities in price $P$ in (a).
(d): The projections $\Braket{\psi^{[\mathrm{maxI}]}|\psi_0}^2$ and
  $\Braket{\psi^{[\mathrm{maxI}]}|\psi_0}^2$ 
are used to determine whether the current 
$I$ is low or high.
The calculations are performed in the price basis 
$Q_k=P^k$  with $n=10$ and $\omega=1$  (entire trading day average).
}
\end{figure*}
Another example to demonstrate\cite{2016arXiv160204423G} the special role of the state $\psi^{[\mathrm{maxI}]}$
can be shown using the basis of $Q_k=P^k$ and $\omega=1$, with the matrices $\Braket{P^j|I|P^k}$
and $\Braket{P^j|P^k}$ in Eq. (\ref{GEV});
these
are now regular volume- and time-  averages over the entire trading day.
In this basis, $\psi^{[\mathrm{maxI}]}(P)$ is a function of trading price $P$.
We calculated the $I_0$, as well as the projections $\Braket{\psi^{[\mathrm{minI}]}|\psi_0}^2$
and  $\Braket{\psi^{[\mathrm{maxI}]}|\psi_0}^2$ which are now the functions of price $P$.
In Fig. \ref{ntPlotChart}, for the same trading day, September 20, 2012, as in Fig. \ref{IHEGCalc}a,
we calculate these values along with the price and price/volume histogram.
In the Fig. \ref{ntPlotChart}b, a histogram of the volume distribution
is presented.\footnote{Price analysis of various kinds, such as skewness, fat tails analysis, etc.,
is often performed by analysts considering charts similar to this price-volume distribution,
but they are not related to market dynamics.}
The price/volume distribution has a maximum around $P=698$.
Whereas most of the shares were actually traded near this price, close to the median, there is no singularity in 
$P(t)$ at this level.
In the Fig. \ref{ntPlotChart}c, the 
execution flow $I_0(P)$ shows a much sharper picture than the one with the volume histogram, Fig. \ref{ntPlotChart}b.
There are high values of $I$ near the  $694$ and $700$ price levels.
These levels correspond to very active trading (large $I$)
around $10$AM and near the market close at $15:30$PM.
There is no singularity in the total volume distribution near these 
$694$ and $700$ price levels because, despite the high values of $I$,
the total time spent at these price levels is low, resulting in low volume traded.
The price, however, exhibits a singularity around 
$694$ and $700$.
This experimentally demonstrates that the price singularity is directly related to the execution flow singularity.
In the Fig. \ref{ntPlotChart}d, the projections
$\Braket{\psi^{[\mathrm{minI}]}|\psi_0}^2$ and $\Braket{\psi^{[\mathrm{maxI}]}|\psi_0}^2$
are presented; these can be used as actionable triggers to identify low or high $I$.

From the two charts -- volume distribution and $I(P)$ --
we can make an important observation about price impact (market impact).
Price impact \cite{2009PhRvE..80f6102M, gatheral2013dynamical, 2014arXiv1412.0141D,bucci2019crossover} is typically considered
as the path-dependent impact of executed shares on asset price.
As we see from the volume distribution chart, the volume near the $694$ price level is rather low,
but the price change and 
$I$ are high. This makes it reasonable to introduce the concept of dynamic impact,
the sensitivity of asset price to the execution rate $I$,
which differs from the regular impact -- the sensitivity of asset price to executed volume.
The charts presented demonstrate that both price and 
execution flow singularities are located at about the same price levels.
A similar situation in time-space can be observed in Fig. \ref{IHEGCalc}.

However, this result was not accurate enough to construct a profitable trading strategy with our available setup.
In this work, we developed a greatly improved directional indicator that brings us close to building such a strategy.
More importantly, the developed methodology allows translating expected changes in execution flow into corresponding price changes.
These new results, however, do not create a ``money-printing'' trading strategy. The current goal is to understand the relation between execution flow and market dynamics, and then to prove this relation experimentally using actual exchange data.

\section{\label{PnLExpressions}P\&L Calculation Methods}

Most trading systems focus on price prediction.
However, a trader is not actually interested in prices; what matters is the P\&L.
From our point of view, the P\&L, not the price, should be the quantity to predict.
Whereas the price $P(t)$ describes the market, the P\&L incorporates both market data and trader actions.
Let us write a formal expression for the calculation of the P\&L of an equity asset.

Define the position change $dS$ -- the number of shares bought ($dS>0$) or sold ($dS<0$) during an interval $dt$.
When integrated over the full time horizon, a trading strategy $dS$ must satisfy
\begin{align}
0=\int dS \label{dSeq0}
\end{align}
This constraint ensures that, for P\&L calculation, the position is closed at the end of the investment horizon.
If a trading strategy is not yet closed at $t_{now}$, one may formally add a single term $-S_0$ for the currently held position:
\begin{align}
S_0&=\int\limits_{-\infty}^{t_{now}} dS
\end{align}
and define the modified trading strategy
\begin{align}
dS^{\prime}&= dS - S_0\delta(t-t_{now})dt  \label{unrealStrategy}
\end{align}
which satisfies (\ref{dSeq0}).
The meaning of this modified strategy is that all held positions are assumed to be sold at $t_{now}$;
if sold at $P^{last}$, this corresponds to the calculation of unrealized P\&L.
For a given strategy $dS$ satisfying (\ref{dSeq0}), its P\&L is
\begin{align}
\mathrm{P\&L}&=-\int P dS
\label{PnLdef}
\end{align}
This is the general form of the P\&L operator.
A simple example: if one buys $v$ shares at $P_1$ and then sells them at $P_2$,
the corresponding $dS/dt=v\delta(t-t_1)-v\delta(t-t_2)$; substituting into (\ref{PnLdef}) gives $\mathrm{P\&L}=v(P_2-P_1)$.
For convenience, it is better to measure $dS$ in the number of shares and use a discrete measure instead of delta functions,
i.e., to consider $dS/dV$ and integrate it over $dV$ in (\ref{dSeq0}) and (\ref{PnLdef}), replacing the integral with a sum.

Integrating (\ref{PnLdef}) by parts, we obtain a different form of the expression, now written in terms of price changes:
\begin{align}
\mathrm{P\&L}&=\int S dP
\label{PnLdefDP} \\
S(t_{start})&=S(t_{end})=0 \label{S0constraint}
\end{align}
The constraints (\ref{S0constraint}) explicitly require that the held position $S(t)$ equals zero
at both the beginning and the end of the trading interval.
This form is often less preferable in practice, since integration over $dP$
is harder to perform than integration over a discrete measure $dS$.
However, it simplifies the consideration of more general forms of trading strategies, such as (\ref{exeFlowTradingStrategySVtau}), and the difficulties in calculating $\mathrm{Tr}\left\| \rho \frac{dP}{dt} \frac{dV}{dt} \right\|$
terms can be overcome using the technique from Appendix \ref{QQMCalculation}.

The P\&L above is presented on a ``cash basis''.
Initially, a trader holds cash and zero asset positions,
trading between them with the goal of ending with zero asset position and a cash position increased by the P\&L.
One can similarly consider a trading process that results in zero cash position and maximal asset position.
In this case, the P\&L is measured in units of asset shares, and all P\&L operator expressions remain the same.
It is also possible to require an explicit percentage split between cash and asset positions
to be achieved at the end of the trading strategy. In this case, the P\&L operator is modified slightly.
In all considerations below, we will use P\&L on a cash basis;
modifications for asset-based P\&L are straightforward.
Although asset-based P\&L may seem unnatural for equities trading, it is commonly used in currency trading.

\begin{figure}[t]
% sed -n '/B\EGIN_PnLExample/,/END_PnLExample/p' ExecutionFlow.tex >/tmp/a.gpl ; gnuplot-wx /tmp/a.gpl
\begin{comment}
# gnuplot commands
#%BEGIN_PnLExample
set output "q.eps"  ;set pointsize 3 ;  l=693
set terminal postscript eps size 12cm,12cm enhanced color font 'Times,28'
  
fn="/home/mal/tmp/pnlPlot_out.csv"
unset grid

set multiplot #layout 3,1
set tmargin 0
set bmargin 0
set lmargin screen 0.05
set rmargin screen 0.99
set xrange [9.85:10.15]
unset xtics
unset ytics
set border 1+2+4+8 lw 1
set yrange [693.5:698]
set size 1,0.491
set origin 0,0.504
set key center left
set label "t" at screen 0.95,0.53
plot fn using ($2/3600e9):($3) with lines lt 1 lc 1 lw 2 title "P"
set yrange [-400:600]
set ytics 100000
set size 1,0.25
set origin 0,0.254 
set key top left
plot fn using ($2/3600e9):($5) with impulses  lt 1 lc 3 lw 6 title "dS", fn using ($2/3600e9):($6) with steps  lt 1 lc 2 lw 2 title "S", fn using ($2/3600e9):(0) with lines lt 1 lc 7 lw 6 notitle
set yrange [-150:720]
set ytics 100000
set size 1,0.25
set origin 0,0.004
set key top left
plot fn using ($2/3600e9):($9) with steps  lt 1 lc 5 lw 2 title "P\\&L", fn using ($2/3600e9):(0) with lines lt 1 lc 7 lw 6  notitle

exit
#%END_PnLExample
\end{comment}

  \includegraphics[width=0.9\columnwidth]{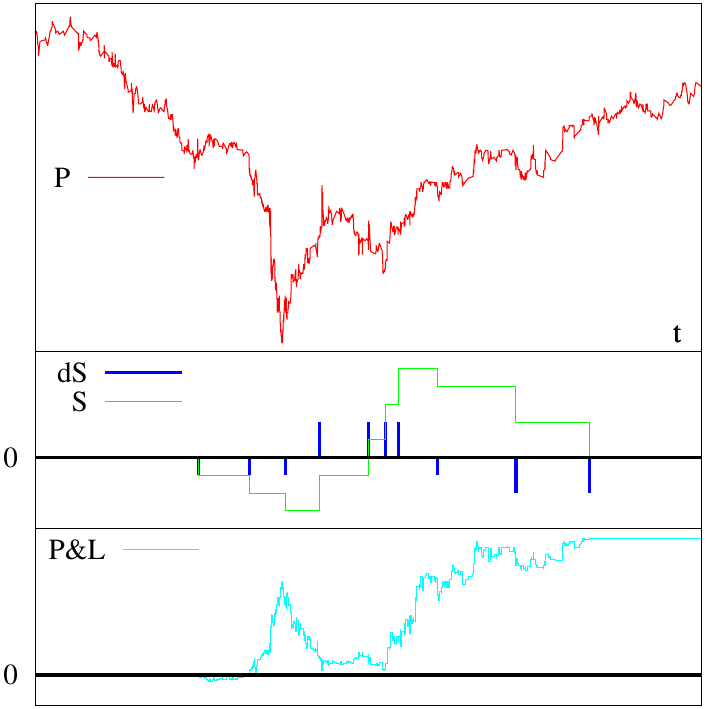}
  \caption{\label{PnLCalc}
A demonstration of P\&L calculation according to (\ref{PnLdef}).
The discrete measure $dS$ represents the trader’s actions, and its integral $S$ gives the position held.
Integrating $dS$ with the asset price yields the P\&L.
It is important to emphasize that the P\&L depends on both the asset price $P(t)$ and the trader’s actions $dS$.
}
\end{figure}

In Fig. \ref{PnLCalc}, we present a simple demonstration of a trading strategy consisting of ten events (blue $dS$ ``impulses'').
The position held is obtained by integrating $dS$, and the P\&L is calculated by integrating $PdS$ (\ref{PnLdef}).
The P\&L depends on both the asset price $P(t)$ and the trader’s actions $dS$.
The ultimate problem of market dynamics is to construct $dS$ from past observations $(t_l, P_l, dV_l)$
such that it consistently yields a positive P\&L.
Consider a few trivial strategies that yield a positive P\&L.

Consider a strategy $S(t)=w(t)dP/dt$, where $w(t)$ is an arbitrary positive function. For simplicity, assume $w=1$,
and that $dP/dt$ is zero on the boundaries of the trading interval, thus the constraints (\ref{S0constraint}) are satisfied.
Substituting this $S(t)$ into (\ref{PnLdefDP}), we immediately obtain a positive P\&L.
Differentiating this $S$, we obtain $dS/dt=d^2P/dt^2$.
This is an important result: the position increment $dS/dt$ should behave as the second derivative of price.
This may look trivial, but it is actually not.
The very important point is the symmetry of the trading strategy’s position increment:
the position increment should have the symmetry of the second derivative of price.
It must change sign for $P\to -P$, and, importantly, must not change sign for $t\to -t$.
Trading strategies that do not exhibit this symmetry will not consistently make money.
There is a well-known mantra in the HFT community: trade the second derivative of price.

Consider a strategy $dS=(P^{F}_{ma}-P)dV$, where $P^{F}_{ma}$ is the ``future'' regular moving average of $\tau$ scale,
calculated on the $[t_{now},t_{now}+\tau]$ interval.
Substituting this $dS$ into (\ref{PnLdef}) yields a positive P\&L proportional to the standard deviation squared.
If using the median price instead of $P^{F}_{ma}$, the strategy is modified
to buy anything below the median price level and sell everything above it.
When using, instead of $P^{F}_{ma}$, the past moving average $P_{ma}$ (calculated on the past $[t_{now}-\tau,t_{now}]$  interval),
we obtain a typical ``mean-reversion'' strategy.
It may perform adequately as long as there is no large market move.
However, when such a move occurs, a catastrophic P\&L loss typically results.

Consider a strategy
$dS=\pm\Big[\lambda^{[\mathrm{maxI}]}{\psi^{[\mathrm{minI}]}}^2(x(t)) - \allowbreak \lambda^{[\mathrm{minI}]}{\psi^{[\mathrm{maxI}]}}^2(x(t))\Big]dV$,
where $\psi^{[i]}$ are the eigenvectors of (\ref{GEV}).
This strategy opens a position at $P^{[\mathrm{minI}]}$ and closes it at $P^{[\mathrm{maxI}]}$.
Whether to go long or short (select the sign of $\pm$) depends on which price is lower.
This serves as an example of a strategy where $dS$ is determined by the probability density calculated from (\ref{GEV}).

These example strategies (along with several others presented in our previous works)
present a self-referential problem: to construct a $dS$ strategy with a positive P\&L, we need to know future prices.
In these examples, we inject future prices into $dS$ to produce a positive P\&L
from the terms $\int PdS$ or $\int SdP$ in the P\&L operator.
Practically, no information about future prices can be used in $dS$.
Yet, to achieve positive P\&L, some information ``from the future'' is required.
As discussed in \cite{2015arXiv151005510G}, prices cannot serve as such a source.
Importantly, any practical $dS$ model must not explicitly depend on asset prices from the future.

However, if we examine the execution flow $I=dV/dt$, we realize that we can have some information
``from the future'' --  specifically, information about the future execution flow.
This implies that a $dS$ model should depend on future execution flow only, not future prices.

\section{\label{impactF}Impact From The Future}
What information about the future can we obtain at $t=t_{now}$
from past observations of the sequence $(t_l, P_l, dV_l)$?
Given the currently observed value of execution flow $I_0 = \Braket{\psi_0|I|\psi_0}$,
we know with certainty that the future execution flow $I_0^{F}$ will be greater than $I_0$,
since additional trading will inevitably occur in the future.
The maximal eigenvalue $\lambda^{[\mathrm{maxI}]}$ of (\ref{GEV}) serves as an estimate of the future execution flow $I_0^{F}$:
\begin{align}
  I_0^{F}&=\lambda^{[\mathrm{maxI}]} \label{iofuture} \\
  dI^{F} &= I_0^{F} -I_0  \label{dI}  \\
  dI^{F} &\ge 0 \label{dIge0}
\end{align}
A very important fact is that the future $I$ estimator, $\lambda^{[\mathrm{maxI}]}$,
is calculated based on already executed trades.
If trading activity ``now'' is slow (i.e., $I_0$ is small),
this indicates that buyers and sellers are not well matched at the current price,
implying that the asset price must adjust.
The price movement is expected to occur due to an increase in future $I$, driven by ``future execution''.
In this sense, the slower the market is now, the more dramatic the expected price movement in the future.
The past most dramatic $I$, represented by $\lambda^{[\mathrm{maxI}]}$,
can therefore serve as a reasonably good estimator (\ref{iofuture}) of the future dramatic $I$.
The eigenvector corresponding to the maximal execution flow is special ---
a kind of ``ground state'', in analogy with quantum mechanics.
Conceptually, this may appear similar to the ``reversion to the moving average'' type of strategy
often applied by market practitioners to asset prices or their standard deviations.
However, this analogy is incorrect.
Experimental observations \cite{2016arXiv160204423G} show that such reasoning
can be applied only to the execution flow $I = dV / dt$,
not to the trading volume, asset price volatility, or any other observable.
Moreover, this prediction works only in one direction --- the execution flow tends to increase.
A criterion for the absence of information about the future can also be formulated:
if the current $I_0$ is close to $\lambda^{[\mathrm{maxI}]}$,
it means that we are already in a ``very dramatic market'' at present,
and thus no additional information about the future state of the market can be inferred:
\begin{align}
    dI^{F}&=0
    \label{dIeq0}
\end{align}
In Fig. \ref{IHEGCalc}, one can identify the ``no information'' moments when $I_0$ (yellow line)
touches $\lambda^{[\mathrm{maxI}]}$ (top pink line).
Similarly, moments of slow current trading activity
(where a dramatic price movement is expected in the future)
can be identified when $I_0$ is close to
$\lambda^{[\mathrm{minI}]}$ (bottom pink line).

The question now is how to use the future $I$ (\ref{iofuture}) to obtain directional price information.
One might formally attempt to add some trading volume at $t=t_{now}$,
as discussed in Section VII.C ``Impact From The Future Operator'' of Ref. \cite{ArxivMalyshkinMuse},
but this approach is likely incorrect, since these trades have not yet occurred.
Instead, the future $I$ should propagate into the dynamic equation through the boundary condition at $t=t_{now}$.

As discussed above, a trader should open a position during liquidity deficits and close it during liquidity excesses.
This statement defines the trading strategy. In the previous section, we developed a method to compute the strategy’s P\&L.
Thus, this liquidity trading strategy can be represented by trading with the following $dS$:
\begin{align}
dS&=dI \label{exeFlowTradingStrategy}
\end{align}
For this trading strategy, the change in position is equal to the change in execution flow.
To calculate its P\&L, one needs to integrate (\ref{exeFlowTradingStrategy}). Over which time interval?
One might think this should be in the $\psi^{[\mathrm{maxI}]}$ state with the measure $d\mu={\psi^{[\mathrm{maxI}]}}^2(x(t))\omega(t)dt$,
but this measure is localized in the past, and the contribution from $t_{now}$, where we know the future $I$,
is small, of order $\Braket{\psi^{[\mathrm{maxI}]}|\psi_0}^2$.
Based on our previous most successful attempt at a directional indicator (\ref{dirOld}),
it is clear that the strategy should be executed over the interval
from the spike in $I$ corresponding to $\lambda^{[\mathrm{maxI}]}$ up to $t_{now}$.
This trading strategy (\ref{exeFlowTradingStrategy}) can be given an alternative interpretation:
let the position $S$ be defined as
\begin{align}
S&=I-\lambda^{[\mathrm{maxI}]} \label{exeFlowTradingStrategyS}
\end{align}
It satisfies the condition (\ref{S0constraint}),
since the strategy starts in the $\psi^{[\mathrm{maxI}]}$ state with $S=0$ (due to the state selection)
and ends at $t_{\mathrm{now}}$ with $S=0$ as a result of the impact from the future (\ref{iofuture});
this form explicitly has zero position when the execution flow $I$ is high.

If the full LOB information is available, a more general form than (\ref{exeFlowTradingStrategyS}) can be introduced.
It may use a measure involving  $t^{\mathrm{in}}$ (\ref{Quadruples}).
For example, with $d\varpi$:
\begin{align}
S&=\frac{dV}{dt}-\lambda^{[\mathrm{max}]}\frac{d\varpi}{dt} \label{exeFlowTradingStrategySVtau}
\end{align}
Here, $\lambda^{[\mathrm{max}]}$ is the maximal eigenvalue of (\ref{GEV}), with
$\Braket{Q_j|\frac{dV}{dt}| Q_k}$ and $\Braket{Q_j|\frac{d\varpi}{dt}| Q_k}$
as the matrices in the left- and right-hand sides;
compare the measure $dV-\lambda^{[\mathrm{max}]}d\varpi$ with the one in footnote \ref{measureFromMinLambda}.
This form has several promising advantages, but it requires $d\varpi$,
which is difficult to obtain for all execution events.
While regulators require the execution volume $dV$ to be provided, the timing $d\varpi$ is not provided.
This approach is typically limited by using a subset of execution data,
obtained from the LOB dynamics of specific exchanges \cite{itchfeed},
allowing the acquisition of (\ref{Quadruples}) data.
In the commonly considered case $d\varpi=dt$, (\ref{exeFlowTradingStrategySVtau})
matches exactly with (\ref{exeFlowTradingStrategyS}) and (\ref{exeFlowTradingStrategy}).
While such a simplified approach works with readily available data, it may have some conceptual limitations.
See Appendix \ref{DirAssym} below for the case when full LOB data is available,
and the execution flow is defined per originating limit order arrival as $dV/d\varpi$ (\ref{imodified}).

When performing the calculations, the situation is simplified by the fact that for the two bases we consider, 
$x=(t-t_{now})/\tau$ and $x=\exp((t-t_{now})/\tau)$
with
$\omega=\exp((t-t_{now})/\tau)$,
both the infinitesimal time shifts and the partial interval integration
preserve the $\omega(t)$ weight and the polynomial basis space.
This means that integration and differentiation can be expressed via the same moments (an analogue of integration by parts).
If there were no $\omega(t)$ weight, this would correspond to plain differentiation and integration operators, but $\omega(t)$ introduces extra terms.
The integration with weight corresponding to ``since $\psi(x)$ until now'' can be obtained via interval partial integration.
This transform is analytically known for the two bases we use, see Appendix A of Ref. \cite{malyshkin2022market}.
Basically, this means that if the value of $f$ in the state $\psi$ is $\Braket{\psi|f|\psi}$,
then the value of $f$ in the state ``since $\psi$ until now'' is $\mathrm{Tr} \rho f$, where
the density matrix $\rho$ is calculated from the polynomial $\psi^2$ as described in Appendix A of Ref. \cite{malyshkin2022market}, also see Appendix \ref{MatrixMeasure} below.
This allows to obtain 
\begin{align}
  f(t_{now})-\Braket{\psi|f|\psi}&=\mathrm{Tr}\left\| \rho \frac{df}{dt}\right\|
  \label{spurFromP}
\end{align}
This is essentially a glorified integration by parts:
the $f$ in the pure state $\Ket{\psi}$ can be expressed via $df/dt$ in the mixed state $\rho$,
which is calculated from $\psi^2$ using an integration-like operation,
see Section II ``Basis Selection'' of Ref. \cite{ArxivMalyshkinMuse}, Section II ``Basic Mathematics'' of Ref. \cite{MalMuseScalp}, and Appendix A of Ref. \cite{malyshkin2022market}.

Having the method (\ref{spurFromP}) to calculate ``since $\psi$ until now'',
let us take $f=I$ and $\psi=\psi^{[\mathrm{maxI}]}$, then calculate the density matrix $\rho$ corresponding to the polynomial $\psi^2(x)$.
We immediately see that if the boundary value $I(t_{now})$ equals the impact from the future (\ref{iofuture}),
we have $0=\mathrm{Tr}\left\| \rho \frac{dI}{dt}\right\|$,
i.e., it satisfies the P\&L constraint (\ref{dSeq0}).
In calculating the P\&L for the liquidity trading strategy (\ref{exeFlowTradingStrategy}),
$dI$ should be used as the position change $dS$ in (\ref{PnLdef}),
and the integral should be replaced by a trace with respect to the density matrix $\rho$.
The P\&L for the trading strategy (\ref{exeFlowTradingStrategy}) provides the directional information.
The algorithm is straightforward:
\begin{itemize}
\item
From past observations, calculate the moments $\Braket{Q_m I}$, construct the matrices $\Braket{Q_j|I| Q_k}$ and $\Braket{Q_j| Q_k}$,
solve the eigenproblem (\ref{GEV}), and determine $\lambda^{[\mathrm{maxI}]}$ and $\psi^{[\mathrm{maxI}]}$.
\item
Using the procedure of Appendix A of Ref. \cite{malyshkin2022market},
construct the density matrix $\rho$ from the polynomial ${\psi^{[\mathrm{maxI}]}}^2(x)$;
$\rho$ corresponds to the state ``since $\psi(x)$ until now''.
\item Calculate the P\&L for the trading strategy (\ref{exeFlowTradingStrategy})
\begin{align}
\mathrm{dir}_{PdI}&=\mathrm{Tr}\left\| \rho \frac{PdI}{dt}\right\| \label{DirOur}
\end{align}
which provides the directional information.
There is no ``$-$'' sign from (\ref{PnLdef}) included in (\ref{DirOur}) to match our old result (\ref{dirOld}).
The same result can be obtained using Eqs. (\ref{exeFlowTradingStrategyS}) and (\ref{PnLdefDP}):
\begin{align}
\mathrm{dir}_{PdI}&=\lambda^{[\mathrm{maxI}]}\left(P^{last}-P^{[\mathrm{maxI}]}\right)
- \mathrm{Tr}\left\| \rho \frac{IdP}{dt}\right\|
\label{pdIexpDtime}
\end{align}
This expression differs from (\ref{DirOur}) only in that
$P^{[\mathrm{maxI}]}=\Braket{\psi^{[\mathrm{maxI}]}|P|\psi^{[\mathrm{maxI}]}}$
is now time-averaged,
rather than volume-averaged ($P^{[\mathrm{maxI}]}$  from (\ref{PIH})) as used in Eqs. (\ref{dirOld}) and (\ref{pdIexp});
for sufficiently large $n$, the two values are nearly identical.

This directional information has a clear meaning:
if the current P\&L of the trading strategy $dS=dI$ (\ref{exeFlowTradingStrategy}) is positive (negative),
then it will \textsl{remain such} for some (rather substantial) time in the future.
A practical application is that when the current $I_0$ is small
(e.g., $\Braket{\psi^{[\mathrm{minI}]}|\psi_0}^2>0.9$)
one should open a long (short) position to capture the future $dI^{F}$ (\ref{dI}).
There is no such information available from a price move:
if the price goes up, it can either continue the trend or bounce back.
The difference between a past price move and the P\&L (\ref{DirOur})
is that the P\&L preserves its sign for a rather substantial period of time.
This is because we determined the optimal time scale of $I=dV/dt$ from (\ref{GEV})
by using $\psi^{[\mathrm{maxI}]}$ to construct the integration measure in (\ref{DirOur}) (density matrix $\rho$).
\end{itemize}
The only remaining difficulty is calculating the matrix elements
$\Braket{Q_j|P\frac{dI}{dt}|Q_k}$
required for taking the trace in (\ref{DirOur}),
an analogue of the P\&L integration (\ref{PnLdef}).
It would be straightforward if the $PdS$ operator were a full differential.
For example, if we formally take the operator $\frac{dPI}{dt}$
as a proxy to $P\frac{dI}{dt}$,
we immediately obtain $\mathrm{dir}=\lambda^{[\mathrm{maxI}]}\left(P^{last}-P^{[\mathrm{maxI}]}\right)$,
which exactly corresponds to our previous result (\ref{dirOld})!
However, this is not a proper liquidity trading strategy
since it introduces an extra term $\frac{dP}{dt}\frac{dV}{dt}$, but it demonstrates the correctness of our approach.
The calculation of the required matrix elements is discussed below in Appendix \ref{QQMCalculation}.
Also see Appendix A of Ref. \cite{malyshkin2022market}.

\section{\label{directionalInfo}Directional Information: A Practical Demonstration}
In this section, we present the directional indicators (\ref{dirOld}) and (\ref{DirOur})
for the same dataset considered above; the datasets from \cite{NYSEtaq} will be discussed later.
The goal of this section is to demonstrate the market microstructure,
especially its directional information.
One might consider processing the data statistically,
but any statistical analysis requires averaging over some time scale,
which would prevent us from examining the market microstructure ---
a system that lacks a characteristic time scale for which stable statistical properties can be obtained
(heteroscedasticity of the market).
The only available source of a time scale is the averaging with the density
matrix $\rho$, obtained from the $\psi^{[\mathrm{maxI}]}$ solution of (\ref{GEV}).
Whereas the market itself does not have a characteristic time scale, market participants do ---
at least the minimal time scale at which they can execute a transaction.
An automated trading machine, built based on the time scale obtained from $\psi^{[\mathrm{maxI}]}$,
also has intrinsic time scales.
They are determined by $\tau$ and the basis dimension $n$.
For the basis $x=\exp((t-t_{now})/\tau)$, $\omega=\exp((t-t_{now})/\tau)$,
the $\Braket{Q_j|I|Q_k}$ matrix has contributions from $\tau/(2n-1)$ to $\tau$.
For the basis $x=(t-t_{now})/\tau$, $\omega=\exp((t-t_{now})/\tau)$,
the $\Braket{Q_j|I|Q_k}$ matrix has contributions from $\tau$ to approximately $2n\tau$.
Whereas a moving average operates with a single time scale,
our approach works with a range of time scales.
The solution $\psi^{[\mathrm{maxI}]}$ corresponds to the optimal one.
In the demonstrations of this section, we use $n=12$ and $\tau=128$s.
The range may not correspond precisely to any specific market,
but the ability to select the proper time scale (from a certain range) is the major result of our work.

\begin{figure}[t]
  \includegraphics[width=0.9\columnwidth]{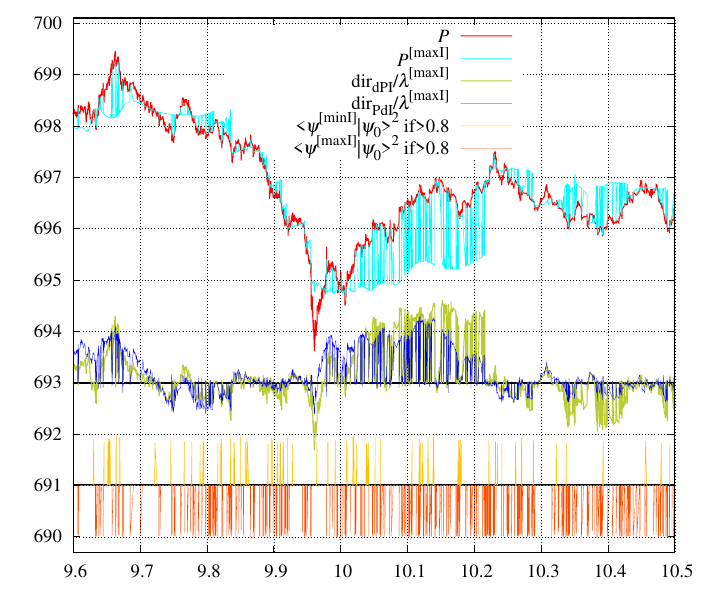}
  \caption{\label{DirectionalO}
The directional information (\ref{dirOld}) and (\ref{DirOur}) (shifted to 693 to fit the chart),
the price, and $P^{[\mathrm{maxI}]}$ (\ref{PIH}) are shown above.
Below (shifted to 691), we present an indicator of low $I$ -- a possible ``entry point'',
$\Braket{\psi^{[\mathrm{minI}]}|\psi_0}^2$ (if $>0.8$),
and an indicator of low $I$ -- a possible ``exit point'',
$\Braket{\psi^{[\mathrm{maxI}]}|\psi_0}^2$ (if $>0.8$),
shown below the 691 level in the plot.
 }
\end{figure}

As discussed above, there should be at least four entry/exit signals (actionable triggers).
In Fig.~\ref{DirectionalO}, we present the directional indicators
$\mathrm{dir}_{dPI}$ (\ref{dirOld}) and $\mathrm{dir}_{PdI}$ (\ref{DirOur}).
One can clearly see that they switch when the market conditions change.
The older indicator $\mathrm{dir}_{dPI}$ \cite{2015arXiv151005510G}, having only a positive measure in $P^{[\mathrm{maxI}]}$  (\ref{PIH}),
represents the difference between the last price and the price in the $\psi^{[\mathrm{maxI}]}$ state.
The indicator  $\mathrm{dir}_{PdI}$ includes an additional term, $\frac{dP}{dt}\frac{dV}{dt}$ (\ref{expressionPdI}),
which provides more ``forward-looking'' information.
Empirical results show that the main concept proposed in \cite{malyshkin2022market} -- 
comparing the terms $I\frac{dP}{dt}$ and $P\frac{dI}{dt}$ -- is not particularly effective.
The best directional indicator is obtained from the $P\frac{dI}{dt}$ term
in the P\&L trading strategy (\ref{exeFlowTradingStrategy}).
Note that this strategy assumes very specific entry/exit levels.
The corresponding entry/exit points are shown on the same chart
as the projections of  $\psi^{[\mathrm{minI}]}$ and $\psi^{[\mathrm{maxI}]}$
on $\psi_0$, exceeding $0.8$. They are marked in orange/red on the chart.

This demonstration shows a highly accurate tracking of directional information.
Of particular interest is the regime switch at $t=9.97$, which is precisely detected by $\mathrm{dir}_{PdI}$ (\ref{DirOur}).
A natural question arises: when does this approach fail?
Typically, this occurs when the basis dimension $n$ and the parameter $\tau$
do not correspond to the actual market dynamics, and the state with the optimal time scale cannot be constructed.
Although not shown in the chart, around $t=14.00$ the trading data from NASDAQ ITCH -- used in all charts above --
become significantly slower (a few thousand transactions every half hour)
compared to the beginning of the trading session (a few thousand transactions every few seconds).
Under such conditions, the chosen value $n=12$ becomes insufficient,
and $\tau=128$s is too small
to construct a $\psi$ corresponding to a large time scale,
and the behavior turns rather random.
A distant analogy would be plotting a moving average with a time window $\tau$ that is too small.
In our case, this corresponds to $\tau$ being so mismatched that the basis of $n$ functions
becomes insufficient to construct the proper state.

\begin{figure}[t]
  \includegraphics[width=0.9\columnwidth]{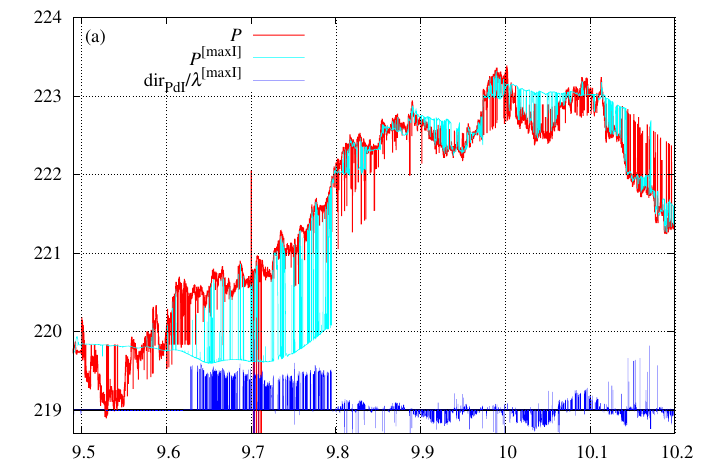}\\
  \includegraphics[width=0.9\columnwidth]{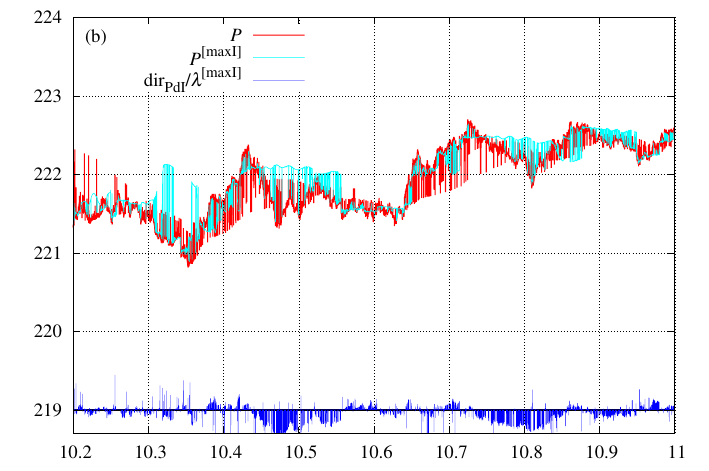}\\
  \includegraphics[width=0.9\columnwidth]{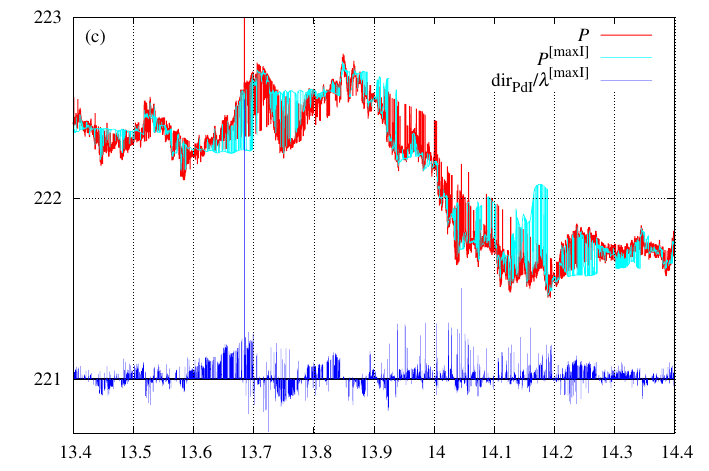}
  \caption{\label{DirectionalTAQ}
The $\mathrm{dir}_{PdI}$ (\ref{DirOur}) is shown for AAPL on 2025.04.01, totaling 594,673 transactions \cite{NYSEtaq}.
The $\mathrm{dir}_{PdI}$ is filtered by entry points; its value is displayed only when
 $\Braket{\psi^{[\mathrm{minI}]}|\psi_0}^2>0.8$,
and otherwise it is set to zero; it is moved to 219 and 221 levels to fit the chart.
$P^{[\mathrm{maxI}]}$ (\ref{PIH}) is also presented.
One can see that the term $\frac{dP}{dt}\frac{dV}{dt}$ (\ref{expressionPdI})
effectively ``removes some signals'' compared to
$\mathrm{dir}_{dPI}=\lambda^{[\mathrm{maxI}]}\left(P^{last}-P^{[\mathrm{maxI}]}\right)$ (\ref{dirOld}).
Periods when the basis dimension $n = 12$ is insufficient for $\tau = 128$s are also observed.
 }
\end{figure}

To demonstrate the approach on appropriate HFT data, we used NYSE TAQ files.
This source contains significantly more transactions than NASDAQ ITCH, making it more suitable for our approach.
See Appendix~\ref{SoftwareDescription} below for a description of software usage.
In Fig.~\ref{DirectionalTAQ}, we present data for AAPL stock on 2025.04.01, totaling 594,673 transactions;
the data is obtained from \cite{NYSEtaq}.
One can see from the figures that the $\psi^{[\mathrm{maxI}]}$
state is actually preserved for a substantial period of time.
This is why the P\&L trading strategy can potentially provide information about the future.
The plots also highlight periods when the basis dimension $n = 12$ is insufficient for $\tau = 128$s.
Based on these market observations, we can conclude the following:
\begin{itemize}
\item Execution flow, $I=dV/dt$, is the driving force of the market;
price singularities are directly observed in Fig.~\ref{IHEGCalc} near large $I_0$,
and especially in Fig.~\ref{ntPlotChart}, where price basis $Q_j=P^j$ is used \cite{2016arXiv160204423G}.
\item The state $\psi^{[\mathrm{maxI}]}$, corresponding to the maximal execution flow solution of (\ref{GEV}),
is relatively stable for a time much longer than the price tick interval.
This stability allows us to extract information based on the impact from the future assumption (\ref{dI}).
\item The method to convert the impact from the future into a possible future price change is the P\&L trading strategy,
$dS = dI$ (\ref{exeFlowTradingStrategy}),
by calculating the P\&L in the state ``since $\psi(x)$ until now'' (\ref{DirOur}).
\end{itemize}

\section{\label{concl}Conclusion}

In this paper, we develop a quantitative approach based on trade execution flow, $I=dV/dt$.
The data typically collected by society consist of individual transactions:
side $A$ sells $v$ units of a good to side $B$ at price $P$, receiving $vP$ dollars.
In each such transaction, supply and demand are perfectly matched.
Information sources where supply and demand are not matched (such as limit order book or advertisement listings)
are much less accessible and collected with far less rigor.
In this work, we develop a dynamic theory that operates solely on transaction data:
instead of stating that price is determined by the balance of supply and demand,
we propose that price is determined by the maximum of the execution flow, $I=dV/dt$,
which can be directly observed from transaction data.

An original mathematical framework, based on the Radon-Nikodym derivative,
is developed to calculate the execution flow from transaction data.
The fundamental question is what information about the future is available to us.
We show that it is information about future execution flow (\ref{iofuture}).
This impact from the future is then converted into the expected price change
using the liquidity trading strategy (\ref{exeFlowTradingStrategy}),
yielding directional information in the form of P\&L (\ref{DirOur}).
A demonstration for a single asset is presented using several data samples.

The theory can be extended to a multi-asset universe. There are two possible approaches:
\begin{itemize}
\item
Consider the capital flow for all assets $a$ of interest, $dC/dt=\sum_a P^{(a)} I^{(a)}$,
and formulate a single eigenvalue problem similar to (\ref{GEV}) for $dC/dt$ instead of $dV/dt$.
\item
Consider each asset separately, applying its own equation (\ref{GEV}) for $I^{(a)}$,
and then combine the results.
\end{itemize}
Our preliminary experiments indicate an advantage of the second approach,
since the states of maximal execution flow for different assets may lead or lag each other in a seemingly random manner.
While a full understanding of multi-asset dynamics remains a subject of future research,
we emphasize that the developed technique for \textsl{incremental} calculation of moments from
the execution flow is highly efficient and capable of processing data in real time.
The directional indicator for an individual asset (\ref{DirOur}) has the dimension of capital traded.
These indicators can be summed across all (or at least the most capitalized) assets to obtain the total market direction:
\begin{align}
\mathrm{DIR}&=\sum_a \mathrm{dir}_{PdI}^{(a)}
\label{DIRTOTAL}
\end{align}
Combined with parallelization of solving the eigenproblem (\ref{GEV}) for each individual asset,
we see no obstacles to deploying this approach in real time across the entire U.S. equity market.
See Appendix \ref{MultiStockCalculations} below for examples and a preliminary analysis.

\begin{acknowledgments}
This research was supported by Autretech Group,
%FMMEM \href{https://xn--80akau1alc.xn--p1ai/}{www.{\fontencoding{T2A}\fontfamily{cmr}\selectfont атретек.рф}},
a resident company of the Skolkovo Technopark.
  We thank our colleagues from the Autretech R\&D department
  who provided insight and expertise that greatly assisted the research.
  Our grateful thanks are also extended
  to Mr. Gennady Belov for his methodological support in doing the data analysis.
\end{acknowledgments}

\appendix
\section{\label{QQMCalculation}Calculation of $\left<Q_j\middle|\frac{dP}{dt}\frac{dV}{dt}\middle|Q_k\right>$
matrix elements from sampled moments}
Direct sampling (\ref{momentsDef}) allows obtaining only the moments of first derivatives.
Second-order derivatives can be obtained either from secondary sampling or from another type of approximation.
The main matrix of interest $\Braket{Q_j|P\frac{dI}{dt}|Q_k}$ can be converted, using integration by parts,
to $\left<Q_j\middle|\frac{dPI}{dt}\middle|Q_k\right>$ (which is trivial to calculate)
and $\left<Q_j\middle|\frac{dP}{dt}\frac{dV}{dt}\middle|Q_k\right>$, which is much more difficult to compute.
In Appendix A of Ref. \cite{malyshkin2022market}, we considered several approximations for calculating the second derivative moments.
The main idea for computing the moments of a product of two functions is to introduce a delta-function-type expression.
\begin{align}
&\Braket{Q_j| f g |Q_k} = \label{deltaFun} \\
&\int\limits_{-\infty}^{t_{now}} \omega(t)dt \int\limits_{-\infty}^{t_{now}}
dt^{\prime} Q_j(x(t)) f(t) \delta(t-t^{\prime})  g(t^{\prime}) Q_k(x(t^{\prime}))
\nonumber
\end{align}
Then change the integration variable to $x$ and use a reproducing kernel
as an approximation of the delta function:
\begin{align}
\mathcal{K}(x,x^{\prime})&=\sum\limits_{j,k=0}^{n_d-1} Q_j(x) G^{-1}_{jk} Q_k(x^{\prime})
\label{approxDelta}
\end{align}
For a fixed $x^{\prime}=x_0$, the reproducing kernel gives a wavefunction localized at $x_0$,
e.g., $\psi_0(x)=const\cdot \mathcal{K}(x,x_0)$, Eq. (\ref{psi0def}), where $const$ is a normalizing constant such that $\Braket{\psi_0|\psi_0}=1$.
If $n_d=n$, then we obtain the familiar approximation for the product of functions\cite{malyshkin2022market}.
\begin{align}
\Braket{Q_j |f g |Q_k}&\approx
\sum\limits_{q,r=0}^{n_d-1} \Braket{Q_j|f|Q_q} G^{-1}_{qr} \Braket{Q_r|g|Q_k}
\label{prodApprox}
\end{align}
This operator approximation, while being non-Hermitian,
creates no problem since it is used only in the calculation of the trace with the Hermitian density matrix $\rho$,
as in (\ref{DirOur}).
Numerical experiments show that it is the moments of $\frac{dP}{dt}\frac{dV}{dt}$
that are well-approximated in this product-type expression.
The moments of functions containing second derivatives (especially of price, e.g.,
$\Braket{Q_jI\frac{d^2P}{dt^2}}$, $\Braket{Q_jV\frac{d^2P}{dt^2}}$, etc.) are particularly poor in this type of approximation.
For simplicity, we will use $f=dP/dt$ and $g=dV/dt$, the moments of which are obtained from sampling (\ref{momentsDef}),
to estimate $\left<Q_j\middle|\frac{dP}{dt}\frac{dV}{dt}\middle|Q_k\right>$.
This is the simplest version of the approximation theory developed in Appendix A of Ref. \cite{malyshkin2022market}.

An important improvement is that now, in the reproducing kernel (\ref{approxDelta}), we take the dimension $n_d>n$.
This creates rectangular $n\times n_d$ matrices
$\left<Q_j\middle|\frac{dP}{dt}\middle|Q_k\right>$ and
$\left<Q_j\middle|\frac{dV}{dt}\middle|Q_k\right>$,
and analytically known Gram matrix (\ref{GramM}) now has dimension $n_d\times n_d$.
Everything else in (\ref{prodApprox}) remains the same; a typical good value for $n_d$ is $n_d\gtrsim 2n$.
The result is a well-approximated matrix
$\left<Q_j\middle|\frac{dP}{dt}\frac{dV}{dt}\middle|Q_k\right>$
of dimension $n\times n$,
which we use to obtain the matrix
$\left<Q_j\middle|P\frac{dI}{dt}\middle|Q_k\right>$
required for P\&L calculation (\ref{DirOur})
of the liquidity trading strategy (\ref{exeFlowTradingStrategy}).
\begin{align}
\left<Q_j\middle|P\frac{dI}{dt}\middle|Q_k\right>&=
\left<Q_j\middle|\frac{dPI}{dt}\middle|Q_k\right>
-
\left<Q_j\middle|\frac{dP}{dt}\frac{dV}{dt}\middle|Q_k\right>
\label{expressionPdI}
\end{align}
If only the first term, $\left<Q_j\middle|\frac{dPI}{dt}\middle|Q_k\right>$, is retained --
then the new result for directional information (\ref{DirOur}) corresponds exactly
to the old result (\ref{dirOld}) obtained in Ref. \cite{2015arXiv151005510G}.
In the general case of (\ref{DirOur}), we have:
\begin{align}
\mathrm{dir}_{PdI}&=\lambda^{[\mathrm{maxI}]}\left(P^{last}-P^{[\mathrm{maxI}]}\right)
- \mathrm{Tr}\left\| \rho \frac{IdP}{dt}\right\| \nonumber \\
&=\mathrm{dir}_{dPI} - \mathrm{Tr}\left\| \rho \frac{IdP}{dt}\right\| \label{pdIexp}
\end{align}

In addition to the expression (\ref{prodApprox}) used to evaluate the product of
$dP/dt$ and $dV/dt$
averaged with some weight $\Pi(x(t))\omega(x(t))dt$,
there is an alternative method involving direct application of (\ref{Ipsi}) followed by numerical integration,
without using the delta-function expansion in (\ref{deltaFun}).
Specifically, substituting the localized state (\ref{psi0def}) with $x_0=y$ into (\ref{Ipsi}),
we obtain the Radon-Nikodym interpolation (\ref{Iy}) of an arbitrary function as a ratio of two polynomials of degree $2n-2$:
\begin{align}
\frac{df}{dt}(y)&=\frac{\sum\limits_{j,j^{\prime},k^{\prime},k=0}^{n-1} Q_j(y) G^{-1}_{jj^{\prime}} \Braket{Q_{j^{\prime}}|\frac{df}{dt}|Q_{k^{\prime}}}
  G^{-1}_{k^{\prime}k} Q_k(y)}
  {
  \sum\limits_{j,k=0}^{n-1} Q_j(y) G^{-1}_{jk} Q_k(y)
  }
  \label{fyinterpolation}
\end{align}
Then, the integral
$\int_{-\infty}^{0}\frac{dP}{dt}\frac{dV}{dt} \Pi(x(t))\omega(x(t))dt$ 
can be estimated using the interpolation expression from (\ref{fyinterpolation}),
yielding an integrand expression involving a product of two ratios of polynomials of degree $2n-2$, the polynomial $\Pi$,
and the weight $\omega$.
This expression can be evaluated numerically using any quadrature method, such as Gaussian quadrature
(with the weight $\omega$, or preferably, with $\Pi\omega$),
or a simple Newton-Cotes formula;
for some $\omega$, the integral can be evaluated analytically.
The obtained integral evaluates $\frac{dP}{dt}\frac{dV}{dt}$
in a pure state if $\Pi=\psi^2$,
or in a mixed state for an arbitrary $\Pi$,
such as $\Pi=J(\psi^2)$, as considered below.
This type of evaluation typically arises when calculating the P\&L in the state determined by Eq. (\ref{GEV})
using a trading model such as (\ref{exeFlowTradingStrategySVtau}).

\subsection{\label{MatrixMeasure}Relation Between Matrix Elements and Measure}
There is an important point to remember when manipulating the matrices directly —
the obtained result may not correspond to any actual measure.
Matrices derived from an observable $f$ do correspond to some measure.
The moments $\Braket{Q_mf}$ of a measure are first obtained through direct sampling (\ref{momentsDef}),
and then the matrix  $\Braket{Q_j|f|Q_k}$ is constructed using the multiplication operator (\ref{multiplicationOperator}).
However, the matrix $\Braket{Q_j|fg|Q_k}$, obtained approximately as a product of matrices using (\ref{prodApprox}),
is generally not Hermitian and may contain approximation artifacts.
The reason is that there do not exist moments $\Braket{Q_mfg}$, $m=0\dots 2n$, corresponding to this matrix.
If these moments $\Braket{Q_mfg}$ were directly sampled,
then the resulting matrix $\Braket{Q_j|fg|Q_k}$ would be Hermitian,
retaining all the essential properties.
The existence of a measure corresponding to the matrix is an important property that must be borne in mind.

In this context, the relation between a pure state $\psi$ and the mixed state $\rho$ ``since $\psi$ untill now''
is important to consider.
We will not repeat all the technical details from Appendix A of Ref.~\cite{malyshkin2022market},
but the idea is straightforward.
An observable $f$ in the pure state $\psi$ is given by $\Braket{\psi^2f}$.
In the state ``since $\psi$ untill now'', we perform a partial integration of $\psi^2$,
and the observable $f$ in this state becomes $\Braket{J(\psi^2)f}$,
where $J(\cdot)$ is a polynomial-to-polynomial mapping that conceptually corresponds to integration by parts.
Technically, this means we have moved from a probability density represented by the polynomial $\psi^2$
to a probability density in the form of a general polynomial $J(\psi^2)$ of the same degree.
The difficulty with the average in the form $\Braket{J(\psi^2) f}$ is that it can only be applied
to observables $f$ for which the moments $\Braket{Q_mf}$ are known.
We therefore seek to construct a density matrix $\rho$ such that
\begin{align}
\Braket{J(\psi^2)f}&=\mathrm{Tr} \rho f \label{tracepf}
\end{align}
is satisfied for an arbitrary $f$.
This problem can be readily solved by constructing the moments of a measure corresponding to averaging with the polynomial
$J(\psi^2)$ (which requires solving a linear system of dimension equal to the number of moments),
and then obtaining the density matrix $\rho$ using the multiplication operator (\ref{multiplicationOperator});
see Appendix A of Ref. \cite{ArxivMalyshkinLebesgue}, ``Density matrix, corresponding to a given polynomial''.

\subsection{\label{DirAssym}Directional Indicator Using Full LOB Information}
As we discussed above, while the execution volume is required by regulators to be provided, the time a limit order spends in the LOB before being matched is available only from full LOB dynamics data.
An execution event (a buy-sell match, which results in a trade) is a match of an initial buy/sell limit order arriving
at $t^{\mathrm{in}}$ (\ref{Quadruples})
to the LOB, to a sell/buy order at time $t$, producing an execution volume $dV$ at time $t$.
When considering the full double auction scheme, the unmatched portion of the orders remains in the LOB and
is the subject of extensive research \cite{bucci2019crossover,gatheral2013dynamical,2014arXiv1412.0141D}. As we discussed above, since 2008--2010, a typical LOB pattern consists of a LOB order being executed almost immediately
(possibly partially) after being placed in the LOB,
with the unmatched portion shortly canceled;
if no execution occurs, the entire limit order is typically canceled almost immediately.

This significantly limits the usefulness of full LOB information in 2026. Let us define the ``modified'' execution flow,
a true execution flow,
$\widetilde{I}$
as the number of matched shares per incoming LOB order time $d\varpi$:
\begin{align}
\widetilde{I}&=\frac{dV}{d\varpi}
\label{imodified}
\end{align}
Whereas the definition in (\ref{IclassicDef}) of $dV/dt$, which we used above,
is defined as the number of shares traded per unit time
of the orders causing the execution, the definition in (\ref{imodified}) is defined per unit
time (\ref{dvpmuSample}) of the initial limit order that initiated the execution.
This allows us to look ``back'' and obtain more fine-grained information about the execution flow.
Also, note that order origination time $t^{\mathrm{in}}$ is much more difficult to manipulate than order volume \cite{2016arXiv160305313G}.

If the origination time $t^{\mathrm{in}}$ (and thus the measure $d\varpi$ (\ref{dvpmuSample}))
is available from the LOB for all execution events, the rest is almost identical to the theory we developed above.
From (\ref{Quadruples}) data, incrementally calculate the moments $\Braket{Q_m\frac{dV}{dt}}$ and $\Braket{Q_m\frac{d\varpi}{dt}}$
using direct sampling (\ref{dVmuSample}) and (\ref{dvpmuSample}) for $dV$ and $d\varpi$, then obtain the matrices
$\Braket{Q_j|\frac{dV}{dt}| Q_k}$ and $\Braket{Q_j|\frac{d\varpi}{dt}| Q_k}$ with (\ref{multiplicationOperator}),
to be used on the left- and right-hand sides of Eq. (\ref{GEV}).
The future value of $\widetilde{I}$ is given by the maximal eigenvalue $\lambda^{[\mathrm{max}]}$,
similarly to (\ref{iofuture}).
\begin{align}
\widetilde{I}_0^{F}&=\lambda^{[\mathrm{max}]}
\label{IFmaxwp}
\end{align}

The trading strategy (\ref{exeFlowTradingStrategySVtau}) is then constructed. The density matrix $\rho$
is obtained from $\psi^{[\max]}$
as discussed above, and P\&L (\ref{PnLdefDP}) is calculated using the technique from Appendix \ref{QQMCalculation}.
\begin{align}
\mathrm{P\&L}&=\mathrm{Tr}\left\| \rho \frac{dP}{dt}\frac{dV}{dt}\right\| -
\lambda^{[\mathrm{max}]}\mathrm{Tr}\left\| \rho \frac{dP}{dt}\frac{d\varpi}{dt}\right\|
\label{TrPL}
\end{align}
When using the form (\ref{PnLdef}) of the P\&L operator, the result is very similar. Applying to
\begin{align}
-\mathrm{P\&L}&=\mathrm{Tr}\left\| \rho P\frac{d^2V}{dt^2}\right\| -
\lambda^{[\mathrm{max}]}\mathrm{Tr}\left\| \rho P\frac{d^2\varpi}{dt^2}\right\|
\label{TrPdS}
\end{align}
integration by parts (\ref{spurFromP}), this gives exactly (\ref{TrPL}) within a term of price
$P$ in $\psi^{[\max]}$ state, averaged with $dV$ and $d\varpi$ measures:
$\Braket{\psi^{[\max]}|P\frac{dV}{dt}|\psi^{[\max]}} -
\lambda^{[\mathrm{max}]}\Braket{\psi^{[\max]}|P\frac{d\varpi}{dt}|\psi^{[\max]}}$.
For a large enough basis size $n$, this difference is close to zero.

This theory can actually be applied to any two measures used in (\ref{imodified}) to define $\widetilde{I}$.
An interesting extension can be the usage of the measures
$dV$ (\ref{dVmuSample}) and $d\widetilde{\varpi}$ (\ref{dvpmuSampleVw})
in the left- and right-hand sides of (\ref{GEV}),
instead of $dV$ and $d\varpi$ for (\ref{imodified}).
With this selection, the eigenvalues of (\ref{GEV}) now have the meaning of the inverse time a limit order spends in the LOB.
The maximal eigenvalue $\lambda^{[\max]}$ now corresponds to the state of the minimal time.
The trading strategy $S$ is determined as we used in (\ref{exeFlowTradingStrategySVtau}).

The dynamic equation, which we formulated in \cite{2015arXiv151005510G,2016arXiv160204423G} in variational form as:
``the future price tends to the value that maximizes the number of shares traded per unit time'',
within this new choice of measures, takes an alternative form:
``the future price tends to the value that, for executed orders, minimizes the time a limit order spends in the LOB before execution''.
Whereas these two forms look similar, they actually present different types of market dynamics.
Both of these forms are compatible with current LOB patterns and use the values that are directly calculated
from the observable sequence (\ref{Quadruples}).
Which of these two formulations is closer to the market dynamics actually observed
in modern markets can be a subject of future research.
However, as we discussed in Section \ref{PnLExpressions} above, the P\&L, not the price change,
should be the quantity to predict. With the two measures in question, $dV$ and $d\widetilde{\varpi}$,
the trading strategy consists of opening a position when the orders that have spent a substantial time in the LOB are executed,
and closing it when the recently added orders are executed.
The position held is determined by Eq. (\ref{exeFlowTradingStrategySVtau}) with the corresponding measures used.
The ultimate market dynamics problem then becomes: which two measures provide a trading strategy $S$ that best determines market dynamics?

For the triple input data $(t_l, P_l, dV_l)$, the only possible option is the execution flow $dV/dt$ (\ref{IclassicDef}),
as shown in all the figures above and Appendix \ref{StateSelectionProblem} below.
When the time $d\varpi_l$,
which represents how long the limit order spent in the LOB before execution,
is available from the data (\ref{Quadruples}), the number of possible options increases.
We have suggested the following pairs:
$(dV, dt)$, $(dV, d\varpi)$,
$(dV, d\widetilde{\varpi})$, and possibly $(d\widetilde{\varpi},dt)$.
The best choice of two measures that best determine market dynamics when the time $d\varpi_l$
is available remains an open question.
Since the time $d\varpi_l$, which represents how long a limit order spent in the LOB, fluctuates greatly,
it is probably poorly suited as an integration measure.

Thus, from general considerations, we can infer that when the origination time $t^{\mathrm{in}}$
is available from market observations (\ref{Quadruples}),
the pair $(d\eta,dt)$ (\ref{dvdvin}) is likely the most suitable candidate.
However, a preliminary analysis conducted in Appendix \ref{EtaFlow} below shows no critical advantage.
Alternatively, the $(dV^{\mathrm{in}},dt)$ (\ref{dvdvin}) pair can possibly be used as an improved version of the $(dV,dt)$.

\section{\label{StateSelectionProblem}Experimental Validation of the Selection of the $\psi^{[\mathrm{maxI}]}$ State}
In Section \ref{impactF}, we selected the state $\psi^{[\mathrm{maxI}]}$ from the eigenproblem (\ref{GEV})
as the state determining market dynamics.
Formally, one could choose a different eigenvalue $\lambda^{[i]}$
in the impact-from-the-future expression (\ref{iofuture});
all calculations could then be repeated using the corresponding alternative eigenvector $\psi$.
The directional indicator (\ref{DirOur}) is the P\&L calculated using the density matrix
$\rho$ obtained from the selected state $\psi$ as in (\ref{tracepf}).
Thus, the selection of the eigenproblem state (\ref{GEV}) is a fundamental decision for constructing the directional indicator.

When the full differential $\frac{dPI}{dt}$ is used in (\ref{DirOur}) instead of
 $\frac{PdI}{dt}$,
only the first term in (\ref{pdIexp}) is obtained.
All possible P\&L outcomes for $\frac{dPI}{dt}$ can then be estimated from an eigenproblem with the matrices
$\Braket{Q_j|(P-P^{last})I|Q_k}$ and $\Braket{Q_j|Q_k}$ on the left- and right-hand sides,
where the eigenvalues provide the possible P\&L outcome values.
We explored this approach several times \cite{2015arXiv151005510G,ArxivMalyshkinMuse},
but without much success; see Ref. \cite{malyshkin2022market}, Section III,
for a table of the different left- and right-hand side matrices we previously considered.
If, on the right-hand side, the matrix $\Braket{Q_j|I|Q_k}$ is used instead of $\Braket{Q_j|Q_k}$,
we essentially obtain all possible realizations of past prices (not P\&L),
producing a plot similar to Fig. \ref{opolMulti}, which corresponds to the specific basis $Q_j = P^j$.

\begin{figure}[t]
  \includegraphics[width=0.9\columnwidth]{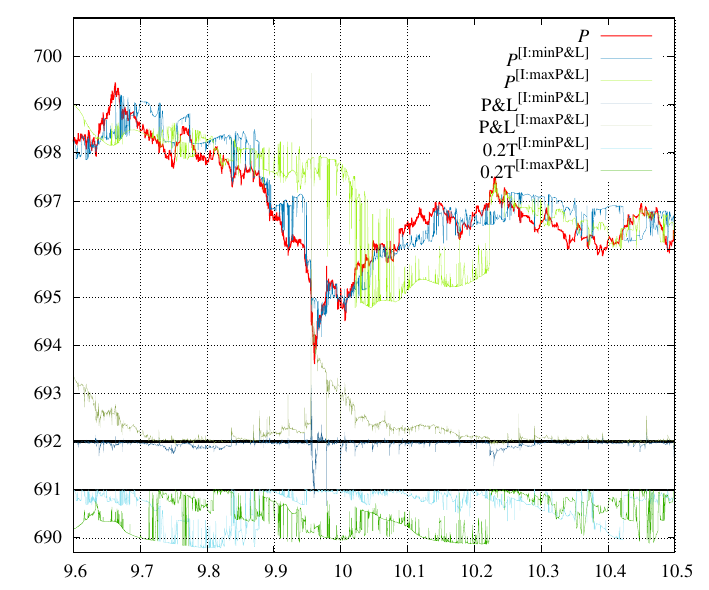}
  \caption{\label{CheckStates}
A demonstration of $P$ (\ref{PIH}), $T$ (\ref{TIH}), and P\&L (\ref{DirOur}) in the states of (\ref{GEV})
producing the minimum (blue palette) and maximum (green palette) values of P\&L (\ref{DirOur}).
The P\&L is multiplied by $5 \cdot 10^{-6}$ and shifted to 692 to fit the chart.
The $T$ values are multiplied by $0.2$ and shifted to 691 for visualization.
The delays are much more pronounced, and ``switching'' behavior is much less clear,
unlike in Figs. \ref{IHEGCalc} and \ref{DirectionalO}.
 }
\end{figure}

This extra term in (\ref{pdIexp}) has important consequences:
first, it makes the effective integration measure non-positive;
second, it breaks the full differential, so that possible P\&L outcomes can no longer be reduced to an eigenproblem
(the range of P\&L outcomes could probably be obtained using
the localized states approach described in Appendix \ref{IKStates} below;
however, we will not consider it in this work).
The effect of this extra term in (\ref{pdIexp}) can be observed in Fig. \ref{DirectionalO},
where $\mathrm{dir}_{PdI}$ (blue) and $\mathrm{dir}_{dPI}$ (olive) are compared.
The extra term reduces $\mathrm{dir}_{dPI}$ by accounting for changes in execution
flow that occurred while price changes were insufficient to affect the indicator.

We now take all eigenvectors $\psi^{[i]}$ of (\ref{GEV}) and, by formally setting
$I_0^{F}=\lambda^{[i]}$ in (\ref{iofuture}) --
which is equivalent to adding the unrealized P\&L term as in (\ref{unrealStrategy}) --
repeat the calculations of $\mathrm{dir}_{PdI}$ from (\ref{DirOur}).
The results are presented in Fig. \ref{CheckStates}, where we examined the other eigenstates of (\ref{GEV}).
The delays are much more pronounced, and the ``switching'' behavior is far less distinct,
unlike in Figs. \ref{IHEGCalc} and \ref{DirectionalO}.
We also considered the states of (\ref{GEV}) filtered by $\lambda^{[i]}\ge I_0$; the results were similar.

These findings lead us to conclude that the state $\psi^{[\mathrm{maxI}]}$ of the eigenproblem (\ref{GEV}),
corresponding to the maximal execution flow $\lambda^{[\mathrm{maxI}]}$, is indeed special ---
a kind of ``ground state'', in analogy with quantum mechanics.
These experiments, involving the additional term in (\ref{pdIexp})
and alternative selections of eigenvectors of (\ref{GEV}),
confirm our earlier conclusion: the state governing market dynamics must not depend on prices.
It is the eigenvector corresponding to the past maximal execution flow that drives market dynamics,
and the past maximal execution flow serves as an estimate of the future flow (\ref{iofuture}).

\section{\label{IKStates}Solving the Optimization Problem in the Localized Basis}

\begin{figure}[t]
% java com/polytechnik/algorithms/TestCall_PnLInPsiHstate --musein_file=dataexamples/aapl_old.csv.gz --musein_cols=9:1:2:3  --n=12 --tau=128 --measure=PnLInPsiHstateLegendreShifted --museout_file=/u1/tmp/pdata/res/PnLInPsiHstate_LegendreShifted_128_12_data1.dat2nd
%
%  sed -n '/B\EGIN_EF_Localized/,/END_EF_Localized/p' ExecutionFlow.tex >/tmp/a.gpl ; gnuplot-wx /tmp/a.gpl
\begin{comment}
# gnuplot commands
#%BEGIN_EF_Localized
set output "q.eps"
set pointsize 3
l=693
set key top right
set key spacing 1.5
set terminal postscript eps size 12cm,10cm enhanced color font 'Times,18'
set xtics 0.05
set ytics 1
set grid front
set xrange [9.9:10.05]
set yrange [691.7:698.3]
set grid front ;s=1e-3;sc=5e-6
fn="/u1/tmp/pdata/res/PnLInPsiHstate_LegendreShifted_128_12_data1.dat2nd"
plot  fn using ($1/3600e9):(l) with lines lc 7 lt 1 lw 2 notitle, fn using ($1/3600e9):($3) with lines lt 1 lc 1 lw 2 title "{/Times-Oblique P}", fn using ($1/3600e9):($28) with lines lt 1 lc 5 lw 4 title "{/Times-Oblique P}^{[maxI]}", fn using ($1/3600e9):($196) with lines lt 1 lc rgb "#BBCD32" lw 1 title "{/Times-Oblique P}@_{loc}^{[maxI]}" , fn using ($1/3600e9):(l-0.2*$26) with lines lt 1 lc rgb "#48cae4" lw 4 title "0.2T^{[maxI]}",  fn using ($1/3600e9):(l-0.2*$197) with lines lt 1 lc 4 lw 1 title "0.2T@_{loc}^{[maxI]}" 
#%END_EF_Localized
\end{comment}

  \includegraphics[width=0.9\columnwidth]{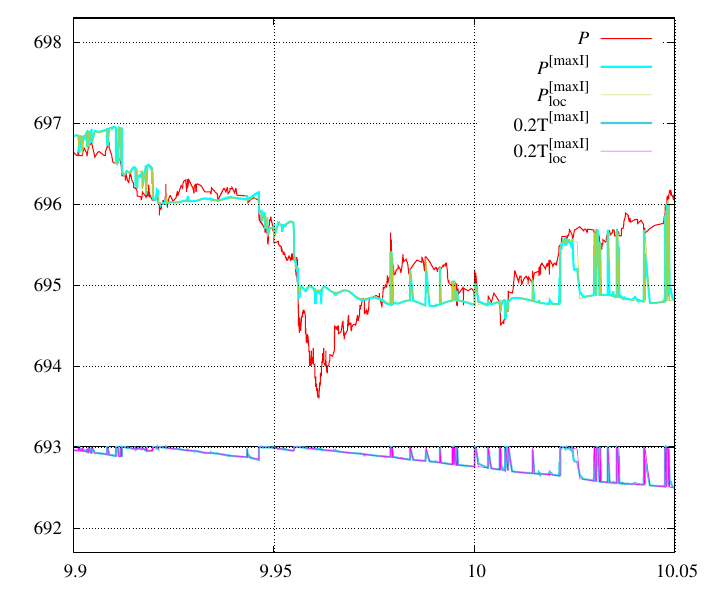}
  \caption{\label{LocStatesPlot}
A presentation of $P^{[\mathrm{maxI}]}$ and $T^{[\mathrm{maxI}]}$ calculated in the state $\psi^{[\mathrm{maxI}]}$
from the solution of (\ref{GEV}) (Fig. \ref{IHEGCalc})
is compared with the results obtained from the localized optimization (\ref{Iy});
the result is obtained for $n=12$ and $\tau=128$s.
One can see very similar results.
This confirms that the $\psi_y(x)$ basis (\ref{psiYlocalized}) can be used for optimization problems
for which an equivalent eigenproblem is not available.
  }
\end{figure}

In the considerations above, we studied the states of maximal execution flow, $I=dV/dt\to\max$,
which led to the eigenproblem (\ref{GEV}).
We may also consider the states related to a large volume traded in the past.
A concept that significantly simplifies this consideration is the Christoffel function:
\begin{align}
K(x)&=\frac{1}{\mathcal{K}(x,x)}=\frac{1}{\sum\limits_{j,k=0}^{n-1} Q_j(x) G^{-1}_{jk} Q_k(x)}
\label{ChristoffelFun}
\end{align}
where $\mathcal{K}(x,x)$ is the reproducing kernel (\ref{approxDelta}),
and $G^{-1}$ is Gram matrix $\Braket{Q_j|Q_k}$ (\ref{GramM}) inverse.
The Christoffel function has been extensively studied in recent works \cite{lasserre2019empirical,lasserre2022disintegration},
it is of significant value for data analysis\cite{bernard2009moments}.
Among the important results of \cite{malyshkin2019radonnikodym}
is the consideration of the Christoffel function spectrum, obtained from the eigenproblem
\begin{align}
\sum\limits_{k=0}^{n-1}\Braket{Q_j|K| Q_k} \alpha^{[i]}_k&=
\lambda^{[i]}\sum\limits_{k=0}^{n-1}\Braket{Q_j| Q_k} \alpha^{[i]}_k \label{GEVChristoffel}
\end{align}
that allows the construction of an invariant expansion --- a promising basis-invariant alternative
to the PCA expansion (which is only unitary-invariant),
a transition from variance expansion to coverage expansion.
It is based on the eigenproblem (\ref{GEVChristoffel}),
where each eigenvector gives the $\lambda^{[i]}$ contribution to coverage,
with the total coverage being $\Braket{1}=\sum_{i=0}^{n-1}\lambda^{[i]}$, see Appendix \ref{ChristoffelSpectrum} below.

Consider the product of execution flow, $I=dV/dt$, with the Christoffel function, $K(x)$.
Extra terms in the denominator make the problem difficult to approach.
However, if we consider only the states localized at $x=y$, denoted as $\psi_y(x)$,
for $y=x_0$ $\psi_y(x)$ is just (\ref{psi0def}),
\begin{align}
\psi_{y}(x)&=\frac{\sum\limits_{i=0}^{n-1}\psi^{[i]}(y)\psi^{[i]}(x)}
           {\sqrt{\sum\limits_{i=0}^{n-1}\left[\psi^{[i]}(y)\right]^2}}
           =
           \frac{\sum\limits_{j,k=0}^{n-1}Q_j(y)G^{-1}_{jk}Q_k(x)}
           {\sqrt{\sum\limits_{j,k=0}^{n-1}Q_j(y)G^{-1}_{jk}Q_k(y)}}             
  \label{psiYlocalized}
\end{align}
In this restricted form of $\psi$, it becomes approachable.
Evaluating an operator in the $\psi_y(x)$ state gives the Radon-Nikodym approximation \cite{malyshkin2019radonnikodym},
which is reduced to a ratio of polynomials of equal degree
\begin{align}
I(y) &\approx \frac{\Braket{\psi_y|I|\psi_y}}{\Braket{\psi_y|\psi_y}} \label{Iy} \\
&=\frac{\sum\limits_{j,j^{\prime},k^{\prime},k=0}^{n-1} Q_j(y) G^{-1}_{jj^{\prime}}\Braket{Q_{j^{\prime}} |I| Q_{k^{\prime}}} G^{-1}_{k^{\prime}k}Q_k(y)}
{\sum\limits_{j,k=0}^{n-1} Q_j(y) G^{-1}_{jk}Q_k(y)}
\nonumber
\end{align}
Compare this expression with the least squares approximation (\ref{Ils}), which is a polynomial.
The $K(y)$ is known analytically from (\ref{ChristoffelFun}), obtain:
\begin{align}
I(y) K(y)&\approx\frac{\sum\limits_{j,j^{\prime},k^{\prime},k=0}^{n-1} Q_j(y) G^{-1}_{jj^{\prime}}\Braket{Q_{j^{\prime}} |I| Q_{k^{\prime}}} G^{-1}_{k^{\prime}k}Q_k(y)}
{\left|\sum\limits_{j,k=0}^{n-1} Q_j(y) G^{-1}_{jk}Q_k(y)\right|^2}
\label{IKapprox}
\end{align}
The product $I(y) K(y)$, calculated using the Radon-Nikodym approximation,
is reduced to a ratio of polynomials.
Contrary to the Rayleigh quotient (\ref{Ipsi}), where the numerator and denominator are of the same degree,
for the product $I(y) K(y)$ the denominator degree, $4n-4$, is twice that of the numerator degree, $2n-2$.
This means we cannot approach the optimization through an eigenvalue formulation. However, by considering polynomials ratio
and using our numerical library \cite{polynomialcode} for manipulating polynomials in an arbitrary basis $Q_j$,
we can find all the zeros of the first derivative of (\ref{IKapprox}) with respect to $y$,
and then select the one corresponding to the maximal $IK$;
in this way, we reduce the optimization problem to finding the polynomial roots (the zeros of the derivative of (\ref{IKapprox})).
The cost of this reduction is that the optimization problem is now formulated in the basis of localized states
(\ref{psiYlocalized}), rather than in the arbitrary basis $\psi$ (\ref{psiEVI}).

\begin{figure}[t]
% java com/polytechnik/algorithms/TestCall_PnLInPsiHstate --musein_file=dataexamples/aapl_old.csv.gz --musein_cols=9:1:2:3  --n=12 --tau=128 --measure=PnLInPsiHstateLegendreShifted --museout_file=/u1/tmp/pdata/res/PnLInPsiHstate_LegendreShifted_128_12_data1.dat2nd
%
%  sed -n '/B\EGIN_EF_IKLocalized/,/END_EF_IKLocalized/p' ExecutionFlow.tex >/tmp/a.gpl ; gnuplot-wx /tmp/a.gpl
\begin{comment}
# gnuplot commands
#%BEGIN_EF_IKLocalized
set output "q.eps"
set pointsize 3
l=693
set key top center
set key spacing 1.5
set terminal postscript eps size 12cm,10cm enhanced color font 'Times,18'
set xtics 0.05
set ytics 1
set grid front
set xrange [9.85:10.15]
set yrange [691.7:698.3]
set grid front ;s=1e-3;sc=5e-6
fn="/u1/tmp/pdata/res/PnLInPsiHstate_LegendreShifted_128_12_data1.dat2nd"
plot  fn using ($1/3600e9):(l) with lines lc 7 lt 1 lw 2 notitle, fn using ($1/3600e9):($3) with lines lt 1 lc 1 lw 2 title "{/Times-Oblique P}", fn using ($1/3600e9):($28) with lines lt 1 lc 5 lw 4 title "{/Times-Oblique P}^{[maxI]}", fn using ($1/3600e9):($202) with lines lt 1 lc rgb "#BBCD32" lw 3 title "{/Times-Oblique P}^{[maxIK]}" , fn using ($1/3600e9):(l-0.2*$26) with lines lt 1 lc rgb "#48cae4" lw 4 title "0.2T^{[maxI]}",  fn using ($1/3600e9):(l-0.2*$203) with lines lt 1 lc 4 lw 3 title "0.2T^{[maxIK]}" 
#%END_EF_IKLocalized
\end{comment}

  \includegraphics[width=0.9\columnwidth]{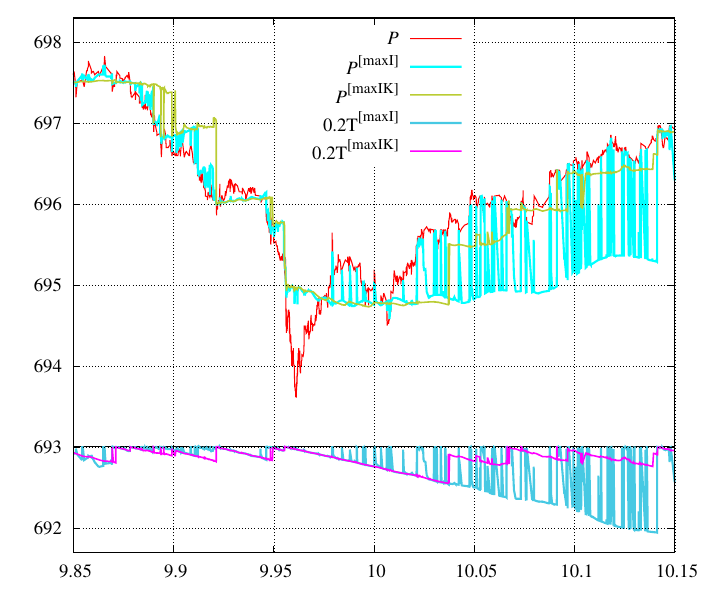}
  \caption{\label{LocIKStatesPlot}
A presentation of $P^{[\mathrm{maxI}]}$ and $T^{[\mathrm{maxI}]}$, calculated in the state $\psi^{[\mathrm{maxI}]}$
that maximizes $I$ (\ref{GEV}), and $P^{[\mathrm{maxIK}]}$ and $T^{[\mathrm{maxIK}]}$, corresponding to a localized $\psi_y$ (\ref{psiYlocalized}) that maximizes $IK$ (\ref{IKapprox}), is shown; the result is obtained for $n=12$ and $\tau=128$s.
Both exhibit state switching, but a switch in the states maximizing $IK$ is less likely.
  }
\end{figure}

Before we consider $IK$, let us compare the two approaches: solve the optimization problem $I \to \max$
in the localized basis (\ref{Iy}), and then compare the result with that obtained from the eigenproblem (\ref{GEV}).
The result is presented in the plot in Fig. \ref{LocStatesPlot}.
One can observe that the eigenproblem (\ref{GEV}) and the localized optimization (\ref{Iy})
produce very similar results for $P$ and $T$.
This allows us to conclude the validity of localized optimization in the basis of $\psi_y(x)$ states (\ref{psiYlocalized}).

Now, having established a technique that takes us beyond the eigenproblem,
let us solve the $IK$ maximization problem (\ref{IKapprox}).
The $IK$ has the meaning of volume, rather than execution flow $I$.
The state $\psi_y(x)$ that maximizes (\ref{IKapprox}) corresponds to the state in which a large trading volume has occurred.
Technically, this is an optimization problem of a ratio of two polynomials.

The result is presented in Fig. \ref{LocIKStatesPlot}.
One can observe a similar type of switching, but the $K(x)$ factor makes the switching less likely,
as it requires a substantial volume to be traded.
The plot demonstrates the validity of the localized states $\psi_y$ (\ref{psiYlocalized}) optimization approach.
Note that this localized optimization is applicable only for one-dimensional problems.
If we were to have a basis of several variables, $Q_j(y)Q_k(z)$,
the optimization (\ref{IKapprox}) would not allow us to find the roots,
whereas the generalized eigenproblem (\ref{GEV}) would still be applicable \cite{malyshkin2019radonnikodym}.

\section{\label{ChristoffelSpectrum}Christoffel Function Coverage Expansion}
The problem (\ref{GEVChristoffel}) can be generalized to a multi-dimensional space to construct a coverage-type expansion.
Consider a sample in an $n$-dimensional space $\mathbf{x}=(x_0,x_1,x_2,\dots,x_{n-1})$;
in the scalar case, we have $x_j = Q_j(x)$.
We also introduce a measure $\Braket{\cdot}$ that enables the calculation of averages $\Braket{x_j|f|x_k}$.
The meaning of this average can be, for example, $\omega dV$, $\omega dt$, or a general sample sum.
The Gram matrix and the Christoffel function are given by:
\begin{align}
G_{jk}&= \Braket{x_j|x_k} \label{GramMD} \\
K(\mathbf{x})&=\frac{1}{\sum\limits_{j,k=0}^{n-1} x_j G^{-1}_{jk} x_k} =
\frac{1}{\sum\limits_{i=0}^{n-1} {\psi^{[i]}}^2(\mathbf{x})}
\label{KMD}
\end{align}
here, $\psi^{[i]}$ is an arbitrary orthogonal basis, satisfying $\Braket{\psi^{[i]}|\psi^{[j]}}=\delta_{ij}$.
Eq.~(\ref{KMD}) is a generalization of (\ref{ChristoffelFun}) to the multi-dimensional space $\mathbf{x}$,
the Christoffel function matrix elements are
\begin{align}
\Braket{x_j|K|x_k}&=\left\langle\frac{x_j x_k}{\sum\limits_{j^{\prime},k^{\prime}=0}^{n-1} x_{j^{\prime}} G^{-1}_{j^{\prime}k^{\prime}} x_{k^{\prime}}}\right\rangle
\label{kMatrixElements}
\end{align}
This requires calculating the average of a ratio of two quadratic functions,
where the one in the denominator is positively definite.
These averages always exist, but their computation is more demanding.
Moreover, due to the presence of the denominator term in (\ref{kMatrixElements}),
they cannot be computed incrementally.
A full scan of the entire sample is typically required to construct the matrix $\Braket{x_j|K|x_k}$.
Consider the eigenproblem
\begin{align}
 \sum\limits_{k=0}^{n-1} \Braket{x_j|K|x_k} \alpha^{[i]}_k &=
  \lambda^{[i]} \sum\limits_{k=0}^{n-1} \Braket{ x_j|x_k} \alpha^{[i]}_k
  \label{GEVKMD} \\
  \psi^{[i]}(\mathbf{x})&=\sum\limits_{j=0}^{n-1} \alpha^{[i]}_j x_j \label{psiMD}
\end{align}
From the definition (\ref{KMD}), it immediately follows that all eigenvalues are positive,
and their sum equals the total measure of the space considered.
\begin{align}
\Braket{1}&=\sum\limits_{i=0}^{n-1} \lambda^{[i]} \label{sumLamMD}
\end{align}
This expansion can be viewed as a generalization of Gaussian quadrature \cite{ArxivMalyshkinLebesgue},
where the weights are
$\lambda^{[i]}$
and the nodes are not discrete measure at $n$ support points, but $n$ probability densities $K{\psi^{[i]}}^2(\mathbf{x})$,
where $\lambda^{[i]}=\Braket{\psi^{[i]}|K|\psi^{[i]}}$.
By sorting the eigenvalues $\lambda^{[i]}$ in descending order, we obtain the factors $\psi^{[i]}(\mathbf{x})$
corresponding to a descending contribution to coverage.
By selecting a few eigenvectors, we can create a projected state that covers a large portion of the observations,
equal to the ratio of the sum of the selected $\lambda^{[i]}$ to the total sum (\ref{sumLamMD}).
This expansion is completely scale-independent,
and the result is invariant under an arbitrary non-degenerate transformation of the $\mathbf{x}$ components:
$x^{\prime}_{j}=\sum_{k=0}^{n-1}T_{jk} x_k$.

For a PCA expansion, we need a function $f$ whose standard deviation we calculate, computing the minimal possible
least squares
\begin{align}
\sigma^2_{\min}&=\Braket{\left(f-\sum\limits_{j=0}^{n-1}\beta_j x_j\right)^2}\to\min \label{LSMinMD}
\end{align}
this is essentially (\ref{LSIproblem}) in the multi-dimensional case.
The standard deviation of $f$ can then be expressed as $\sigma^2_{\min}=\Braket{(f-\overline{f})^2}-\sum_{i=0}^{n-1}\sigma^2_{i}$,
where the contributions $\sigma^2_{i}$ correspond to the eigenvectors of an eigenproblem derived from (\ref{LSMinMD}),
obtained by performing an eigen-decomposition of the covariance matrix and expanding $f$ in the resulting eigenbasis.
Selecting a few of the largest contributions yields the PCA factors ``explanation'' of $f$.
This expansion, however, is only unitary invariant (e.g., the solution will change if we rescale one of the $x_k$),
and it requires the introduction of some function $f$, the variation of which is expanded.
In contrast, the coverage expansion (\ref{sumLamMD}) requires no function $f$
and directly selects the states with the maximal probability of occurrence.
This expansion is of great value for the problem of clustering,
where selecting a few most probable states is of critical importance \cite{malyshkin2019radonnikodym}.

In some situations, when the behavior of $f$ needs to be inferred from the behavior of $\frac{df}{dt}$,
it is convenient to consider the matrix elements
of the same structure as in (\ref{kMatrixElements}):
the average of a ratio of two quadratic functions, where the one in the denominator is positively definite.
Similar to the calculation in (\ref{measuresList}), taking the matrix elements of $df/dt$
replaces the summation over $t_{l}-t_{l-1}$ with a summation over $f_{l}-f_{l-1}$.
The expression for $\Braket{\cdot}$ is identical to (\ref{momentsDef}),
except that, instead of a polynomial $Q_j(x(t_l))$, we now have a ratio of two quadratic functions on $x_m^{(l)}$:
\begin{align}
\Braket{x_j|K\frac{df}{dt}|x_k}&=\left\langle\frac{x_j x_k \frac{df}{dt}}{\sum\limits_{j^{\prime},k^{\prime}=0}^{n-1} x_{j^{\prime}} G^{-1}_{j^{\prime}k^{\prime}} x_{k^{\prime}}}\right\rangle
\label{kMatrixElementsF}
\end{align}
Then we solve a generalized eigenproblem with the matrices $\Braket{x_j|K\frac{df}{dt}|x_k}$ and $\Braket{x_j|x_k}$.
This approach is analogous to the treatment of $K\frac{dV}{dt}$ discussed in Appendix \ref{IKStates} above.

A trivial example. Let $f$ being some portfolio value, and $\frac{df}{dt}$ being daily portfolio change,
$x_j$ are the factors affecting the porfolio value,
and the measure $\Braket{\cdot}$ is taken as a sum over the days $l=1\dots l_{last}$, $l_{last}\gg n$, with $\omega^{(l)}=1$:
\begin{align}
\Braket{x_j|K\frac{df}{dt}|x_k} &=
\sum\limits_{l=1}^{l_{last}} \frac{x^{(l)}_j x^{(l)}_k }{\sum\limits_{j^{\prime},k^{\prime}=0}^{n-1} x^{(l)}_{j^{\prime}} G^{-1}_{j^{\prime}k^{\prime}} x^{(l)}_{k^{\prime}}} (f_l -f_{l-1})
\label{KdfMatrixElements}
\end{align}
To calculate these matrix elements, the sample $x^{(l)}_m,f_l$ must be scanned twice:
first to build the Gram matrix $G_{jk}$, and second to compute the matrix elements in (\ref{KdfMatrixElements}).
This differs from the polynomial moment calculation, where all moments can be obtained
in a single scan using the expansion (\ref{NewtonBinomial}), for example,
in the standard deviation calculation (\ref{sigmaMovaverage}).
The eigenproblem
\begin{align}
\left|K\frac{df}{dt}\middle|\psi^{[i]}\right>&=\lambda^{[i]}\left|G\middle|\psi^{[i]}\right> \label{GEVdKf}
\end{align}
allows us to decompose the P\&L contributions by individual factors.
The sum of all eigenvalues $\lambda^{[i]}$ equals the total change in the portfolio value over the entire period,
\begin{align}
\sum\limits_{i=0}^{n-1}\lambda^{[i]}&=\Braket{\frac{df}{dt}}=\sum\limits_{l=1}^{l_{last}} f_l -f_{l-1}=f_{l_{last}}-f_0
\label{sumDfEqdiff}
\end{align}
compare this expression with (\ref{sumLamMD}).
The solution of (\ref{GEVdKf}) can also be interpreted as a form of Lebesgue quadrature,
where the weights $\lambda^{[i]}$ represent P\&L contributions (not necessarily positive),
and the nodes are not discrete measure at $n$ support points but rather $n$ probability densities  $K{\psi^{[i]}}^2(\mathbf{x})$,
where $\lambda^{[i]}=\Braket{\psi^{[i]}|K\frac{df}{dt}|\psi^{[i]}}$.
Using the expression (\ref{KMD}) for $K(\mathbf{x})$ in an orthogonal basis $\psi^{[i]}$,
we obtain the explicit expression for the probability density:
\begin{align}
K{\psi^{[i]}}^2&=\frac{{\psi^{[i]}}^2}
{\sum\limits_{j=0}^{n-1} {\psi^{[j]}}^2}
\label{WexpressionExplicit}
\end{align}
This yields (\ref{sumLamMD}) and provides the probabilistic interpretation of the factors in the eigenproblem (\ref{GEVdKf})
with respect to the total P\&L in (\ref{sumDfEqdiff}).
For other forms of Lebesgue quadrature, see \cite{ArxivMalyshkinLebesgue}.
Note that the observable (total P\&L) is obtained as a sum of eigenvalues (Lebesgue weights),
representing a form of density matrix average,
rather than as a sum of eigenvalues multiplied by squared projections, as in traditional PCA.

If the Christoffel function $K$ is not used on the left-hand side --
i.e., if we consider an eigenproblem with the matrices $\Braket{x_j|\frac{df}{dt}|x_k}$ and $\Braket{x_j|x_k}$ --
then the $\lambda^{[i]}$ would describe contributions to daily returns, rather than to the total P\&L.
This situation is similar to that considered in Eq. (\ref{GEV}) for calculating the execution flow.
It is the presence of $K$ that allows the eigenvalues to describe contributions to the total P\&L (rather than to daily changes),
which is a significant advantage for risk analysis.

A final important note on the eigenproblem (\ref{GEVdKf}) is that, contrary to PCA --
where all eigenvalues are positive and represent contributions to the variance of $f$ --
in (\ref{GEVdKf}), the eigenvalues $\lambda^{[i]}$ represent the corresponding P\&L contributions (\ref{sumDfEqdiff}) (which may be negative),
while the eigenvectors correspond to contributions to the probability distribution with density (\ref{WexpressionExplicit}).
This allows the expansion (\ref{GEVdKf}) to separately analyze asymmetric P\&L factors with positive and negative contributions.

\section{\label{SoftwareDescription}Software Usage Description}

The software \cite{polynomialcode} is written in Java.
The codebase is fairly large, but all code within the package \texttt{\seqsplit{com/polytechnik/trading/}} --
which constitutes the largest part of it -- represents our earlier,
less successful attempts and has since been converted into unit tests.
To test the provided software, install Java 25 or later.
Download the source code \cite{polynomialcode} from the archive
\href{http://www.ioffe.ru/LNEPS/malyshkin/AMuseOfCashFlowAndLiquidityDeficit.zip}{\texttt{\seqsplit{AMuseOfCashFlowAndLiquidityDeficit.zip}}},
then decompress and recompile it:
\begin{verbatim}
unzip AMuseOfCashFlowAndLiquidityDeficit.zip
javac -g com/polytechnik/*/*java
\end{verbatim}
Then run the software using the sample data located in the \texttt{\seqsplit{dataexamples/}} directory.
Here, we use the backslash ``$\backslash$'' to split lines to fit the two-column PRE format;
BASH interprets it correctly, allowing the commands to be copied directly from the article into the BASH prompt.
\begin{verbatim}
java com/polytechnik/algorithms/\
TestCall_PdIver2 \
 --musein_file=dataexamples/aapl_old.csv.gz \
 --musein_cols=9:1:2:3 \
 --n=12 \
 --tau=128 \
 --measure=CommonlyUsedMomentsLegendreShifted \
 --museout_file=/tmp/museout_PdIver2_128_12.dat
\end{verbatim}
and 
\begin{verbatim}
java com/polytechnik/algorithms/\
TestCall_PdIver2 \
 --musein_file=dataexamples/\
taq_AAPL_20250401.csv.gz \
 --musein_cols=4:1:2:3 \
 --n=12 \
 --tau=128 \
 --measure=CommonlyUsedMomentsLegendreShifted \
 --museout_file=/tmp/mo_PdIver2_128_12_taq.dat
\end{verbatim}
The file specified with \texttt{\seqsplit{--museout\_file=}} contains the results.
The two generated files above include most of the results presented in this paper
and are obtained solely from data in the \texttt{\seqsplit{dataexamples/}} directory.
For a general file from NYSE TAQ \cite{NYSEtaq}, one needs to create a \verb+.csv+
file to use as input for \texttt{\seqsplit{--musein\_file=}}.
Original daily TAQ files from NYSE are typically not time-sorted; to create a time-sorted file, run:
\begin{verbatim}
com/polytechnik/taq/sort_taq_file.sh orig_TAQ.gz
\end{verbatim}
The script \texttt{\seqsplit{sort\_taq\_file.sh}} sorts the TAQ records chronologically.
The script may need to be edited to adjust the temporary directory,
as the generated files are large and a temp directory of over 10Gb is required.
The name of the generated file is printed to \texttt{\seqsplit{stdout}} upon script completion.
The resulting \textsl{sorted} file (we recommend compressing and renaming it to \texttt{\seqsplit{sorted\_NYSE\_TAQ\_file.gz}})
contains all TAQ transactions in chronological order. These ``sorted'' files,
converted from the original TAQ data, can be downloaded from
\url{https://mega.nz/folder/uORjRboa\#bnNJnMt0bQRMkgLvhf5Xuw}.
Next, the data must be filtered to extract only execution transactions for the required stocks. To do this, run:
\begin{verbatim}
java com/polytechnik/taq/\
TAQPrintOutput\$DumpTickersExe \
 sorted_NYSE_TAQ_file.gz \
 >/tmp/all_NYSE_TAQ.csv 2>/tmp/diag.cap
\end{verbatim}
This script generates the file \texttt{\seqsplit{all\_NYSE\_TAQ.csv}}
containing (ticker,time,price,shares) data (\ref{inpData4Columns}),
which can be used with the code presented in this paper.
The file \texttt{\seqsplit{diag.cap}} contains stock trading volumes and traded capital;
it is required to select the instruments of interest and to verify that the calculated volumes
match those reported for that day, e.g., by \href{https://finance.yahoo.com/}{Yahoo Finance}.
If the output needs to be filtered for specific stocks, such as AAPL, add a stock filter list after the input filename.
\begin{verbatim}
java com/polytechnik/taq/\
TAQPrintOutput\$DumpTickersExe \
 sorted_NYSE_TAQ_file.gz AAPL \
 >/tmp/AAPL_NYSE_TAQ.csv 2>/tmp/diag.cap
\end{verbatim}
The resulting four-column file (\ref{inpData4Columns}), \texttt{\seqsplit{AAPL\_NYSE\_TAQ.csv}},
can be used as demonstrated above.
It can be \verb+gzip+-compressed for convenience. For some selected assets, pre-generated files are available at
\url{https://mega.nz/folder/uORjRboa\#bnNJnMt0bQRMkgLvhf5Xuw}.
Thus, the conversion software of NYSE TAQ data to \verb+.csv+ format (\ref{inpData4Columns})
is tested for the latest version,
\href{https://www.nyse.com/market-data/historical/daily-taq}{TAQ v4.2}.

The creation of \verb+.csv+ files from the NASDAQ ITCH feed \cite{itchfeed}
is described in Appendix A of Ref. \cite{2016arXiv160204423G}.
Currently, only ITCH 4.1 is implemented; conversion for ITCH 5.0 is straightforward but has not yet been completed.

\subsection{\label{MultiStockCalculations}Computation of Multi-Asset Indicators}
The computation of multi-asset indicators is considerably more demanding in terms of computational resources.
Such computations can be classified into two categories: local and non-local.
In the local computation approach, the statistical moments for each asset are computed independently,
and the subsequent quantities (such as eigenproblem solutions) are obtained solely from the corresponding asset’s data.
The aggregate market-level indicator is then derived by combining these individual results.
An example of such an indicator is $\mathrm{DIR}$ from (\ref{DIRTOTAL}).
A key advantage of this approach is that computations for different assets can be efficiently parallelized.
In contrast, the non-local computation approach involves first calculating the moments for each asset
and then performing a combined computation that couples data across assets.
An even more non-local case arises when cross-asset data are required already at the stage of moment calculation.
Although such computations are difficult to parallelize,
they may offer computational benefits when the total number of operations is small --
for example, when a single eigenproblem must be solved instead of one for each asset.

The implementation of the directional indicator $\mathrm{DIR}$ from (\ref{DIRTOTAL})
is available in both parallel and single-threaded versions.
An example of the parallel implementation is shown below:
\begin{verbatim}
java com/polytechnik/algorithms/\
TestCall_MultiStockParallelPdIver2 \
 --musein_file=taq_TSLA_..._20250401.csv.gz \
 --assets=TSLA:SPY:NVDA:QQQ:AAPL:AMZN:\
MSFT:META:PLTR:IWM:TQQQ:JNJ \
 --musein_cols=4:0:1:2:3 \
 --n=12 \
 --tau=128 \
 --measure=CommonlyUsedMomentsLegendreShifted \
 --museout_file=/tmp/taq20250401_128_P12_...dat
\end{verbatim}
It processes the four-column \verb+.csv+ file (\ref{inpData4Columns}), generated for
twelve high-capitalization assets
\texttt{\seqsplit{TSLA:SPY:NVDA:QQQ:AAPL:AMZN:MSFT:META:PLTR:IWM:TQQQ:JNJ}},
from the original daily TAQ data, as described above.
Several preprocessed files of this type are available for download from
\url{https://mega.nz/folder/uORjRboa\#bnNJnMt0bQRMkgLvhf5Xuw}.

The program outputs the aggregated $\mathrm{DIR}$ from (\ref{DIRTOTAL}) based on the asset directional indicators
$\mathrm{dir}_{PdI}$ from (\ref{DirOur}), the older result $\mathrm{dir}_{dPI}$ from (\ref{dirOld}),
as well as $\mathrm{dir}_{PdI}$ computed using the extended basis ($n_d>n$) in (\ref{prodApprox}).
All outputs are aggregated over the assets specified via the \verb+--assets=+ option.
The scale of $\mathrm{DIR}$ is also output and can be estimated as 
\begin{align}
\mathrm{SCALE}&=\sum_a P^{last\, (a)}\lambda^{[\mathrm{maxI}]\,(a)}
\label{SCALEDIR}
\end{align}
For a single asset, $\mathrm{DIR}/\mathrm{SCALE}$ roughly corresponds to the relative price change.
The computed data for individual assets are also saved to the same file,
allowing comparison with results obtained using the
\texttt{\seqsplit{com/polytechnik/algorithms/TestCall\_PdIver2.java}}
driver described above.

\begin{figure}[t]

\includegraphics[width=0.9\columnwidth]{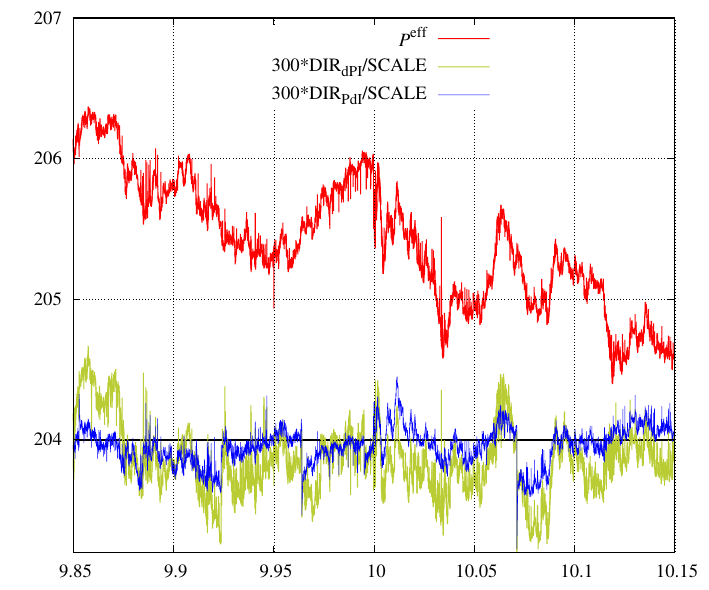}
  \caption{\label{MultiStockPlot}
The multi-asset indicator (\ref{DIRTOTAL}) was calculated for the 12 most heavily capitalized stocks on April 1, 2025:
\texttt{TSLA SPY NVDA QQQ AAPL AMZN MSFT META PLTR IWM TQQQ JNJ}.
Both forms,
$\mathrm{DIR}_{PdI}$ (\ref{dirOld}) and $\mathrm{DIR}_{dPI}$ (\ref{pdIexp}),
are presented. The scale (\ref{SCALEDIR}) is used to normalize them.
The result is multiplied by 300 and shifted up to 204 to fit the chart.
Overall, the results are similar to those for a single asset, as shown in Fig. \ref{DirectionalO}.
  }
\end{figure}

The implementation
\texttt{\seqsplit{com/polytechnik/algorithms/TestCall\_MultiStockPdIver2.java}}
is identical to
\texttt{\seqsplit{com/polytechnik/algorithms/TestCall\_MultiStockParallelPdIver2.java}}
but runs in single-thread mode, typically being much slower when aggregating data for a substantial number of assets.
The implementation processes every tick from (\ref{inpData4Columns}) for the selected assets.
For the asset corresponding to the current tick, the moments are updated with the new (time, price, shares) event.
For all other selected assets, the time reference is shifted to the current $t_{now}$,
so that all assets share the same $t_{now}$ in the same basis.
For each asset,
the eigenproblem (\ref{GEV}) is then solved, directional indicators are obtained,
and the total market direction  $\mathrm{DIR}$ (\ref{DIRTOTAL})
is calculated by aggregating the individual asset directional indicators.
Note that, since we have individual asset matrices $\Braket{Q_j|f|Q_k}$
for any observable $f$, we can calculate, for example,
asset cross-correlations using the matrix multiplication approximation (\ref{prodApprox})
without performing simultaneous sampling of two assets.
We employed similar techniques in our previous works (see e.g. Appendix E of \cite{2015arXiv151005510G}):
cross-correlations were obtained accurately,
but they were less successful in predicting market direction.
In the present work, the aggregated flow $\mathrm{DIR}$ (\ref{DIRTOTAL}) is calculated,
providing the most reliable directional signal we have obtained so far.

A typical result is presented in Fig. \ref{MultiStockPlot}.
We selected the 12 most heavily capitalized stocks on April 1, 2025:
\texttt{TSLA SPY NVDA QQQ AAPL AMZN MSFT META PLTR IWM TQQQ JNJ}.
Then, for every tick, we solved the eigenproblem (\ref{GEV})
for each stock and computed the indicator (\ref{DIRTOTAL}) in both the
$\mathrm{DIR}_{PdI}$ (\ref{dirOld}) and $\mathrm{DIR}_{dPI}$ (\ref{pdIexp})
versions. The full trading day -- over 8 million execution events for these 12 stocks --
was processed in about an hour on very modest hardware.
The application of the obtained directional information requires advanced multi-stock analysis;
here, we present only a simple indicator.
We construct an index-like quantity, $P^{eff}$, to assess the overall market direction:
\begin{align}
P^{eff}&=\frac{\sum\limits_a P^{(a)} V_d^{(a)}}{\sum\limits_a V_d^{(a)}}
\label{indexPeff}
\end{align}
where $V_d^{(a)}$ is the daily trading volume (for April 1, 2025) for asset $a$.
For each tick, the values of $\mathrm{DIR}_{PdI}$ and $\mathrm{DIR}_{dPI}$
are computed as the sum of 12 eigenproblem solutions, one eigenproblem solved independently for each individual asset.
The results in Fig. \ref{MultiStockPlot} are similar to those for a single asset in Fig. \ref{DirectionalO}.
At price singularities, the value of $\mathrm{DIR}_{dPI}$  is often roughly twice as large as that of $\mathrm{DIR}_{PdI}$,
a relationship previously observed in \cite{malyshkin2022market} when comparing the $PdI$ and $IdP$ terms.
The full differential $dPI$ corresponds to (\ref{dirOld}),
with the only source of reduced delay being the ``switching'' of the selected eigenstate in (\ref{GEV}).
The $PdI$ expression (\ref{pdIexp}) contains an additional term which,
on the one hand, yields a more ``second-derivative–like'' type of information,
but at the same time reduces the magnitude by approximately a factor of two.

In this appendix, we demonstrated that the proposed technique is applicable to multi-asset analysis,
but the specifics of its application are highly problem-dependent and will be addressed elsewhere.

\begin{figure}[t]

\includegraphics[width=0.9\columnwidth]{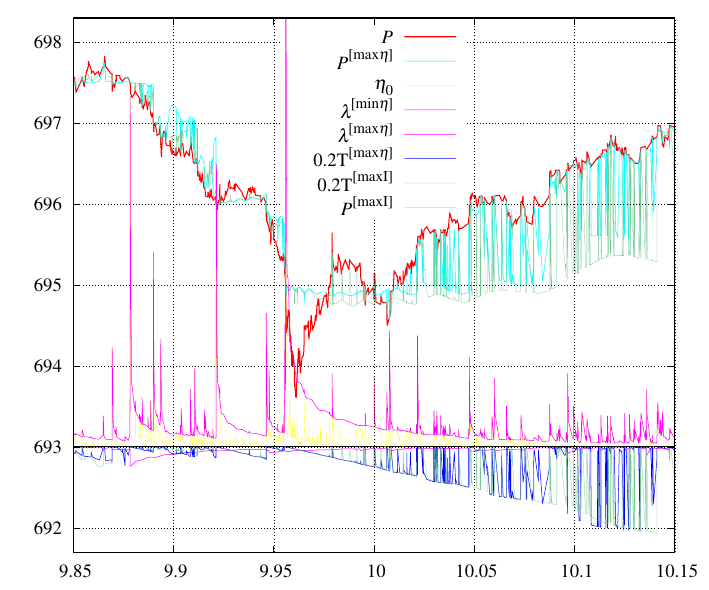}
  \caption{\label{IHEta}
The states of the maximal  $d\eta/dt$ (\ref{dvdvin}),
for the same data sample as in Fig. \ref{IHEGCalc}a, are considered.
Overall, the ``time to now'' (\ref{TIH}) and price in this state (\ref{PIH}) for $\psi^{[\max\eta]}$
are similar to those in the state $\psi^{[\max I]}$, which we previously considered in Fig. \ref{IHEGCalc}a.
These are shown in SeaGreen color in this figure.
The state with the minimal $d\eta/dt$ does not exhibit prominent market dynamics.
  }
\end{figure}

\section{\label{EtaFlow}Demonstrating the In/Out LOB Execution Flow: $d\eta/dt$}
Whereas most of this paper dealt with the absolute execution flow $dV/dt$ (\ref{IclassicDef}),
it may be of interest to consider the in/out flow. For the available full LOB data,
$d\eta/dt$ (\ref{dvdvin})
is the first candidate to consider.
By taking into account the time the initiating LOB order arrived,
we can construct two measures difference $d\eta/dt$. The developed theory can then be applied to it.
As discussed earlier, the first step should be to consider the most prominent states,
those corresponding to maximal/minimal eigenvalues (\ref{GEV}), with the matrices
$\Braket{Q_j|\frac{d\eta}{dt}| Q_k}$ and $\Braket{Q_j| Q_k}$ 
on the left and right-hand sides.
For the state with the maximal eigenvalue, the results
(for the same data as in Fig. \ref{IHEGCalc}a) are presented in Fig. \ref{IHEta}.
Overall, they are similar. Note that $\eta_0=\Braket{\psi_0|\frac{d\eta}{dt}|\psi_0}$
is always positive since, for trades executed ``now'', only LOB removal can occur, as added trades are not yet executed.
The negative $\lambda^{[\min\eta]}$
corresponds to the states localized in the past, where the orders that were later executed were initially added.
This state does not exhibit prominent dynamics, so it was not included in Fig. \ref{IHEta}.

The qualitative behavior is better observed when considering the ``time to now'' distance (\ref{TIH}),
as it fluctuates less than price (\ref{PIH}).
In Fig. \ref{IHEta}, the blue line corresponding to the state $\psi^{[\max\eta]}$
is similar to the SeaGreen lines of the state $\psi^{[\max I]}$.
A similar situation is observed with the price value in this state (light blue vs SeaGreen).

The class
\texttt{\seqsplit{com/polytechnik/freemoney/LOBorigTimeMeasure.java}}
provides the implementation of $d\eta/dt$ dynamics
with unit test that can be run as
\begin{verbatim}
java com/polytechnik/trading/\
TestLOBorigTimeMeasure
\end{verbatim}
and the result itself can be generated as
\begin{verbatim}
java com/polytechnik/algorithms/\
TestCall_LOBorigTimeMeasure \
 --musein_file=dataexamples/aapl_old.csv.gz \
 --musein_cols=9:1:2:3:6 \
 --n=12 \
 --tau=128 \
 --measure=\
CommonlyUsedMomentsWithOrigLegendreShifted \
 --museout_file=/tmp/museout_LOBorig_128_12.dat
\end{verbatim}

\bibliography{LD,mla,econ}

@MISC{polynomialcode,
author={Malyshkin, Vladislav Gennadievich},
title={{The code for polynomials calculation}},
note={\url{http://www.ioffe.ru/LNEPS/malyshkin/code.html}
and an \href{https://disk.yandex.ru/d/AtPJ4a8copmZJ?locale=en}{alternative location}.
},
year="2014",
url={http://www.ioffe.ru/LNEPS/malyshkin/code.html}
}

@article{2015arXiv151005510G,
  title={{Mathematical Foundations of Realtime Equity Trading. Liquidity Deficit and Market Dynamics. Automated Trading Machines}},
  author={Malyshkin, Vladislav Gennadievich and Bakhramov, Ray},
  journal={arXiv preprint arXiv:1510.05510},
  year={2015},
  url={http://arxiv.org/abs/1510.05510},
  doi={10.48550/arXiv.1510.05510}
}

@TECHREPORT{itchfeed, 
   author       = {{Nasdaq OMX}}, 
   type         = {Report}, 
   year         = 2014,
   institution  = "Nasdaq OMX", 
   title        = {{NASDAQ TotalView-ITCH 4.1}},
   url = {http://www.nasdaqtrader.com/content/technicalsupport/specifications/dataproducts/nqtv-itch-v4_1.pdf},
   note={{see sample data files at \url{https://emi.nasdaq.com/ITCH/}
and newest version specification \href{https://www.nasdaqtrader.com/content/technicalsupport/specifications/dataproducts/NQTVITCHSpecification.pdf}{TotalView-ITCH 5.0}
}}
}

@TECHREPORT{NYSEtaq,
 author       = {{NYSE}}, 
   type         = {Report}, 
   year         = 2025,
   institution  = "NYSE", 
   title        = {{Daily TAQ Client Spec v4.2}},
   url={https://www.nyse.com/market-data/historical/daily-taq},
   note={
   see sample data files at
   \url{https://ftp.nyse.com/Historical\%20Data\%20Samples/DAILY\%20TAQ/},
   which provide two days of free data every quarter
   }
}

@MISC{nasdaqord,
author = {Nikolaus Hautsch and Ruihong Huang},
title = {Limit Order Flow, Market
Impact and Optimal Order
Sizes: Evidence from NASDAQ
TotalView-ITCH Data
},
year = "2011",
institution  = "Institute for Statistics and Econometrics and Center for Applied Statistics and Economics (CASE)",
url={http://sfb649.wiwi.hu-berlin.de/papers/pdf/SFB649DP2011-056.pdf}
}

@ARTICLE{totik,
   author       = {Totik, Vilmos }, 
   journal      = {Surveys in Approximation Theory},
   title        = {{Orthogonal Polynomials}}, 
   year = "11~" # nov # "~2005",
   volume       = "1", 
   pages        = "70-125",
   doi = {10.48550/arXiv.math/0512424},
   url={https://arxiv.org/pdf/math/0512424}
}

@ARTICLE{2009PhRvE..80f6102M,
   author = {{Moro}, E. and {Vicente}, J. and {Moyano}, L.~G. and {Gerig}, A. and 
	{Farmer}, J.~D. and {Vaglica}, G. and {Lillo}, F. and {Mantegna}, R.~N.
	},
    title = "{Market impact and trading profile of hidden orders in stock markets}",
  journal = {\pre},
archivePrefix = "arXiv",
   eprint = {0908.0202},
 primaryClass = "q-fin.TR",
     year = 2009,
    month = dec,
   volume = 80,
   number = 6,
      eid = {066102},
    pages = {066102},
      doi = {10.1103/PhysRevE.80.066102},
   url = {http://arxiv.org/abs/0908.0202}
}

@ARTICLE{2014arXiv1412.0141D,
   author = {{Donier}, J. and {Bonart}, J. and {Mastromatteo}, I. and {Bouchaud}, J.-P.
	},
    title = "{A fully consistent, minimal model for non-linear market impact}",
  journal = {ArXiv e-prints},
archivePrefix = "arXiv",
   eprint = {1412.0141},
 primaryClass = "q-fin.TR",
     year = 2014,
    month = nov,
   url = {http://arxiv.org/abs/1412.0141},
   doi={10.48550/arXiv.1412.0141}
}

@article{gatheral2013dynamical,
  title={Dynamical models of market impact and algorithms for order execution},
  author={Gatheral, Jim and Schied, Alexander},
  journal={HANDBOOK ON SYSTEMIC RISK, Jean-Pierre Fouque, Joseph A. Langsam, eds},
  pages={579--599},
  year={2013},
  url={http://ssrn.com/abstract=2034178}
}

@ARTICLE{2016arXiv160204423G,
   author = {Malyshkin, Vladislav Gennadievich},
    title = {{Market Dynamics. On Supply and Demand Concepts}},
  journal = {ArXiv e-prints},
archivePrefix = "arXiv",
   eprint = {1602.04423},
     year = 2016,
    month = feb,
    note={\url{http://arxiv.org/abs/1602.04423}},
   url = {http://arxiv.org/abs/1602.04423}
}

@ARTICLE{ArxivMalyshkinMuse,
 author = {Malyshkin, Vladislav Gennadievich},
    title = {{Market Dynamics. On A Muse Of Cash Flow And Liquidity Deficit}},
  journal = {ArXiv e-prints},
archivePrefix = "arXiv",
   eprint = {1709.06759},
  primaryClass = "q-fin.TR",
     year = 2017,
    month = sep,
   url = {https://arxiv.org/abs/1709.06759},
   doi={10.48550/arXiv.1709.06759}
}

@article{2016arXiv160305313G,
  title={{Market Dynamics vs. Statistics: Limit Order Book Example}},
  author={Malyshkin, Vladislav Gennadievich and Bakhramov, Ray},
  journal = {ArXiv e-prints},
archivePrefix = "arXiv",
   eprint = {1603.05313},
 primaryClass = "q-fin.TR",
     year = 2016,
    month = mar,
    url={https://arxiv.org/abs/1603.05313},
    doi={10.48550/arXiv.1603.05313}
}

@article{ArxivMalyshkinLebesgue,
  title={{On Lebesgue Integral Quadrature}},
  author={Malyshkin, Vladislav Gennadievich},
  journal={arXiv preprint arXiv:1807.06007},
  year={2018},
  url={https://arxiv.org/abs/1807.06007},
  doi={10.48550/arXiv.1807.06007}
}

@article{MalMuseScalp,
  title={{Market Dynamics: On Directional Information Derived From (Time, Execution Price, Shares Traded) Transaction Sequences}},
  author={Malyshkin, Vladislav Gennadievich},
  journal={arXiv preprint arXiv:1903.11530},
  year={2019},
  url = {https://arxiv.org/abs/1903.11530},
  doi ={10.48550/arXiv.1903.11530}
}

@article{lasserre2019empirical,
  title={{The empirical Christoffel function with applications in data analysis}},
  author={Lasserre, Jean-Bernard and Pauwels, Edouard},
  journal={Advances in Computational Mathematics},
  pages={1--30},
  year={2019},
  publisher={Springer},
  doi={10.1007/s10444-019-09673-1},
  url={https://arxiv.org/pdf/1701.02886}
}

@article{malyshkin2019radonnikodym,
  title={{On The Radon-Nikodym Spectral Approach With Optimal Clustering}},
  author={Malyshkin, Vladislav Gennadievich},
  journal={arXiv preprint arXiv:1906.00460},
  year={2019},
  url={https://arxiv.org/abs/1906.00460},
  doi={10.48550/arXiv.1906.00460}
}

@book{bernard2009moments,
  title={{Moments, positive polynomials and their applications}},
  author={Lasserre, Jean-Bernard},
  volume={1},
  year={2009},
  publisher={World Scientific},
  url={https://www.worldscientific.com/worldscibooks/10.1142/p665},
  doi={10.1142/p665}
}

@article{malyshkin2022market,
  title={{Market Directional Information Derived From (Time, Execution Price, Shares Traded) Sequence of Transactions. On The Impact From The Future}},
  author={Malyshkin, Vladislav Gennadievich and Belov, Mikhail Gennadievich},
  journal={arXiv preprint arXiv:2210.04223},
  year={2022},
  doi={10.48550/arXiv.2210.04223}
}

@article{lasserre2022disintegration,
  title={{A disintegration of the Christoffel function}},
  author={Lasserre, Jean B},
  journal={Comptes Rendus. Math{\'e}matique},
  volume={360},
  number={G9},
  pages={1071--1079},
  year={2022},
  doi={10.5802/crmath.380}
}

@MISC{FinancialData,
author={},
title={{Financial data}},
note={
The financial data we used is available for download from
\url{https://disk.yandex.ru/d/AtPJ4a8copmZJ} for NASDAQ
and from \url{https://mega.nz/folder/uORjRboa\#bnNJnMt0bQRMkgLvhf5Xuw} for NYSE TAQ.
These links should also be available from the personal page
\url{http://www.ioffe.ru/LNEPS/malyshkin/code.html}
},
year="2025",
url={http://www.ioffe.ru/LNEPS/malyshkin/code.html}
}

@article{polanyi1957aristotle,
  title={{Aristotle discovers the economy}},
  author={Polanyi, Karl},
  journal={Trade and market in the early empires},
  pages={64--94},
  year={1957},
  publisher={The Free Press New York},
  url={https://archive.org/details/in.gov.ignca.36501}
}

@article{bucci2019crossover,
  title={{Crossover from linear to square-root market impact}},
  author={Bucci, Fr{\'e}d{\'e}ric and Benzaquen, Michael and Lillo, Fabrizio and Bouchaud, Jean-Philippe},
  journal={Physical review letters},
  volume={122},
  number={10},
  pages={108302},
  year={2019},
  publisher={APS},
  doi={10.1103/PhysRevLett.122.108302}
}

@article{kearns2003penn,
  title={{The Penn-Lehman automated trading project}},
  author={Kearns, Michael and Ortiz, Luis},
  journal={IEEE Intelligent systems},
  volume={18},
  number={6},
  pages={22--31},
  year={2003},
  publisher={IEEE},
  doi={10.1109/MIS.2003.1249166}
}

@article{lebaron2006agent,
  title={{Agent-based computational finance}},
  author={LeBaron, Blake},
  journal={Handbook of computational economics},
  volume={2},
  pages={1187--1233},
  year={2006},
  publisher={Elsevier},
  doi={10.1016/S1574-0021(05)02024-1}
}

@article{chakole2021q,
  title={{A Q-learning agent for automated trading in equity stock markets}},
  author={Chakole, Jagdish Bhagwan and Kolhe, Mugdha S and Mahapurush, Grishma D and Yadav, Anushka and Kurhekar, Manish P},
  journal={Expert Systems with Applications},
  volume={163},
  pages={113761},
  year={2021},
  publisher={Elsevier},
  doi={10.1016/j.eswa.2020.113761}
}

@article{donier2016walrasPAPER,
  title={{From Walras’ auctioneer to continuous time double auctions: A general dynamic theory of supply and demand}},
  author={Donier, Jonathan and Bouchaud, Jean-Philippe},
  journal={Journal of Statistical Mechanics: Theory and Experiment},
  volume={2016},
  number={12},
  pages={123406},
  year={2016},
  publisher={IOP Publishing},
  doi={10.1088/1742-5468/aa4e8e}
}

@book{walras2013elementsBOOK,
  title={{Elements of pure economics: Or the theory of social wealth}},
  author={Walras, Leon},
  year={2013},
  publisher={Routledge},
  doi={10.4324/9781315888958},
  url={https://api.taylorfrancis.com/content/books/mono/download?identifierName=doi&identifierValue=10.4324/9781315888958&type=googlepdf}
}

\end{document}